\documentclass[a4paper,12pt]{article}
\pdfoutput=1
\usepackage{geometry}
\usepackage{amsmath,amssymb,amsfonts}
\usepackage{xcolor,graphicx,cite,soul}
\usepackage{array,booktabs,longtable}
\usepackage{caption,microtype}
\usepackage{subcaption}
\usepackage{hyperref}
\usepackage{appendix}

\newcommand{\lsim}
{\;\raisebox{-.3em}{$\stackrel{\displaystyle <}{\sim}$}\;}
\newcommand{\gsim}
{\;\raisebox{-.3em}{$\stackrel{\displaystyle >}{\sim}$}\;}

\newcommand\R{R_\mathrm{2HDM}}

\newcommand\al{\alpha}
\newcommand\be{\beta}
\newcommand\tb{\tan\beta}

\newcommand\CBA{c_{\beta - \alpha}}
\newcommand\SBA{s_{\beta - \alpha}}

\newcommand\ReDiag{\mathop{%
  \raise .5pt\hbox{[}%
  \widetilde{\mathrm{Re}}%
  \raise .5pt\hbox{]}}}
\newcommand\ReOffDiag{\mathop{%
  \raise .5pt\hbox{$\llbracket$}%
  \widetilde{\mathrm{Re}}%
  \raise .5pt\hbox{$\rrbracket$}}}

\newcommand\MW{m_W}
\newcommand\MZ{m_Z}
\newcommand\Mh{m_h}
\newcommand\MH{m_H}
\newcommand\MA{m_A}
\newcommand\MHp{m_{H^\pm}}

\newcommand\msq{m_{12}^{2}}

\newcommand\refeq[1]{Eq.~(\ref{#1})}
\newcommand\refeqs[1]{Eqs.~(\ref{#1})}
\newcommand\refta[1]{Tab.~\ref{#1}}
\newcommand\refse[1]{Sect.~\ref{#1}}

\newcommand\citere[1]{Ref.~\cite{#1}}
\newcommand\citeres[1]{Refs.~\cite{#1}}

\newcommand{\CP}{{\cal CP}}
\newcommand{\cp}{{\CP}}

\newcommand{\tev}{\,\, \mathrm{TeV}}
\newcommand{\gev}{\,\, \mathrm{GeV}}

\newcommand{\br}{\text{BR}}

\def\order#1{\ensuremath{{\cal O}(#1)}}
\def\reffi#1{\mbox{Fig.~\ref{#1}}}

\def\Ga{\Gamma}

\def\la{\lambda}

\newcommand{\lahHpHm}{\ensuremath{\la_{hH^+H^-}}}

\definecolor{Orange}{named}{orange}
\definecolor{Purple}{named}{purple}
\definecolor{Lightblue}{cmyk}{0.9,0.1,0.1,0.3}
\definecolor{dgelborange}{cmyk}{0.,0.3,0.5, 0.}
\definecolor{Lila}{rgb}{0.5,0.,1}
\definecolor{Darkgreen}{rgb}{0.,.7,0.2}

\graphicspath{{figs/}}
\allowdisplaybreaks
\begin{document}
\thispagestyle{empty}

\def\thefootnote{\fnsymbol{footnote}}

\begin{flushright}
\mbox{}
IFT-UAM/CSIC-23-52
\end{flushright}

\vspace{0.5cm}

\begin{center}

{\large\sc 
{\bf Role of \boldmath{$\lambda_{hH^+H^-}$} in Higgs boson decays $h$ to $bs$ in the 2HDM}}

\vspace{1cm}

{\sc
F.~Arco$^{1,2}$%
\footnote{email: Francisco.Arco@uam.es}%
, S.~Heinemeyer$^{2}$%
\footnote{email: Sven.Heinemeyer@cern.ch}%
~and M.J.~Herrero$^{1,2}$%
\footnote{email: Maria.Herrero@uam.es}%
}

\vspace*{.7cm}

{\sl
$^1$Departamento de F\'isica Te\'orica, 
Universidad Aut\'onoma de Madrid, \\ 
Cantoblanco, 28049, Madrid, Spain

\vspace*{0.1cm}

$^2$Instituto de F\'isica Te\'orica (UAM/CSIC), 
Universidad Aut\'onoma de Madrid, \\ 
Cantoblanco, 28049, Madrid, Spain

}

\end{center}

\vspace*{0.1cm}

\begin{abstract}
\noindent
Within the Two Higgs Doublet Model (2HDM) with $\cp$ conservation and 
a softly broken $Z_2$ symmetry, we analyze the flavor changing Higgs 
decays $h\to bs$ ($bs$ refers jointly to the two decay channels $b\bar s$ and $\bar b s$),  where $h$ is identified with the SM-like Higgs boson 
discovered at the LHC. 
We provide a comprehensive study of the decay width $\Gamma (h \to bs)$ with particular  focus on the most relevant effects from the triple Higgs coupling $\lambda_{hH^+H^-}$.
Furthermore, we consider all the relevant theoretical and experimental constraints to determine which predictions 
for the $\br\left(h\to bs\right)$ are still allowed by the current data.
We find that the predictions for $\br\left(h\to bs\right)$ in types II and III can be several orders of magnitude smaller compared to the SM value.  
In contrast,  in type I and IV we find that the predicted 
enhancements in the decay rates with respect to the SM of up to about 70\% and 50\%,  respectively,  are still allowed.
We discuss how these deviations from the SM are caused by interference effects controlled by the coupling $\lambda_{hH^+H^-}$ which can be large for very heavy $H^\pm$.  
To better understand the role of $\lambda_{hH^+H^-}$ in the 
$ h \to bs$ decay we derive and analyze here the analytical results for the $hbs$ one-loop effective vertex that is generated by integrating out the heavy $H^\pm$.
\end{abstract}

\def\thefootnote{\arabic{footnote}}
\setcounter{page}{0}
\setcounter{footnote}{0}

\newpage


\section{Introduction}
\label{sec:intro}

After the discovery of the Higgs boson 
\cite{ATLAS:2012yve,CMS:2012qbp} one of the main goals is to measure its properties with the highest possible accuracy.
The current measurements, within the experimental and theoretical uncertainties, are compatible with the predictions for the Higgs boson in the Standard Model (SM).
Nevertheless,  the triple and quartic Higgs self-couplings still remain unmeasured,  leaving plenty of room for new physics in the Higgs-boson sector.
In particular, flavor-changing Higgs interactions can provide an interesting scenario to search for new physics.
In the SM, flavor-changing neutral currents (FCNC) only occur at the loop level, and are thus highly suppressed. 
Therefore, physics beyond the SM (BSM) could predict much larger FCNC compared to the SM. 
One example are the decays $h\to b\bar s$ and $h\to \bar b s$, which we denote together as $h\to bs$.
The first studies of the SM loop-induced $hbs$ coupling date from the 80s~\cite{Willey:1982mc,Grzadkowski:1983yp},  and more recent works, with the current data, report a prediction of $\br\left(h\to bs\right)\sim10^{-7}$ within the SM~\cite{Benitez-Guzman:2015ana,Aranda:2020tqw}.
This value is very far from the expected experimental sensitivity in the short-medium term~\cite{Blankenburg:2012ex,Barducci:2017ioq}.

One of the most investigated models with Higgs-extended sectors is the Two Higgs Doublet Model (2HDM)~\cite{Gunion:1989we,Aoki:2009ha,Branco:2011iw}. 
The $\cp$-conserving 2HDM predicts five physical Higgs bosons: $h$ and $H$ are $\cp$-even, $A$ is $\cp$-odd, and $H^\pm$ are two charged Higgs bosons, with the masses denoted as $\Mh$, $\MH$, $\MA$ and $\MHp$, respectively.
We will assume that the lightest $\cp$-even Higgs boson $h$ is the SM-like boson discovered with a mass of $\sim 125\gev$.
The addition of the second Higgs doublet can induce FCNC at the tree level, which are highly constrained by the experimental data.
It is usual to consider a $Z_2$ symmetry in the 2HDM to avoid them, which leads to four different Yukawa textures~\cite{Glashow:1976nt,Aoki:2009ha}, providing the so-called four 2HDM types.
This symmetry can be softly broken by the parameter $m_{12}^2$,  with dimension of mass squared.
Nevertheless, the 2HDM scalar sector can induce new sources of FCNC interactions at the loop level in addition to those present in the SM. At one loop, the charged Higgs boson $H^\pm$ mediates these FCNC interactions, which can also involve couplings among the 2HDM scalar states. 

In the case of the $h\to bs$ decay,  the triple Higgs coupling $\lahHpHm$ enters relevantly the 2HDM prediction.  
With our convention, the Feynman rule for the $hH^+H^-$ interaction is given by $-iv\lahHpHm$. 
Therefore, the triple Higgs coupling $\lahHpHm$  will be crucial in the 2HDM prediction, as it can additionally and significantly contribute to the flavor-changing neutral current rates. 
The 2HDM prediction for $\br\left(h\to bs\right)$ without tree-level FCNC was analyzed in \citere{Arhrib:2004xu}.
This decay was also studied in the 2HDM with an arbitrary Yukawa structure in \citeres{Botella:2014ska,Crivellin:2017upt,Herrero-Garcia:2019mcy}.
These 2HDM models\footnote{They are sometimes referred in the literature as 2HDM type III, not to be confused with the 2HDM Yukawa type III, see below.}  induce FCNC at the tree level and require some fine-tuning in the Yukawa parameters to avoid other constraints from flavor observables, and thus we will not consider them here.
The goal of the present paper is to update the 2HDM prediction from \citere{Arhrib:2004xu} considering the current known properties of the SM-like Higgs boson $h$, as well as the exclusion limits from the BSM Higgs-boson searches.
Furthermore, we will emphasize chiefly the role of $\lahHpHm$ in the prediction of $\br(h \to bs)$.
From previous works, it is known that $\lahHpHm$ can be as large as ${\cal O}(30)$ in all 2HDM types while respecting all relevant theoretical and current experimental constraints~\cite{Arco:2022xum,Arco:2020ucn}. 
The 2HDM prediction for $h\to bs$ can exhibit a strong constructive or destructive interference effect, depending on the sign of $\lahHpHm$ and the 2HDM Yukawa type. 

In this work, we will provide a comprehensive analysis of
the effects from $\lahHpHm$, possibly leading to large distortions with respect to the SM prediction, while still in agreement with the current constraints to the 2HDM.
To analyze in more detail the role of $\lahHpHm$ in the $h\to bs$  decay, we will also present here a computation of the effective $hbs$~vertex based on the full one-loop calculation and assuming a heavy $H^\pm$.
This study is done by means of an expansion of the one-loop decay amplitude in inverse powers of the large mass $\MHp$. This will allow us to
analyze in detail the most prominent
effects from the very heavy charged Higgs in the flavor changing decay rates.
It is found that they can yield (depending on the choice of the other 2HDM parameters) a non-vanishing constant (with $\MHp$)  shift with respect to the SM rates in the large $\MHp$ limit.  The effective vertex presented here,  therefore,  summarizes the leading (possibly non-decoupling) effects from the heavy charged Higgs boson within the 2HDM. This effective vertex is calculated for all four 2HDM types.

The structure of this article is as follows.
In \refse{sec:2hdm} we briefly review the 2HDM and fix our notation.
In \refse{sec:Gamma} we discuss the 2HDM prediction for the $h\to bs$ partial width and we analyze the role of $\lahHpHm$ in the prediction.
We first focus on the simplified scenario in the alignment limit in \refse{sec:Gamma-align}.
We also provide an effective vertex description of the 2HDM effects in this decay in the alignment limit after the performance of a large $\MHp$ series expansion in \refse{sec:largeMHp}.
In \refse{sec:Gamma-noalign} we turn to the study outside the alignment limit.
In \refse{sec:BR} we consider all relevant theoretical and experimental constraints for the 2HDM, using a similar analysis as in \cite{Arco:2022xum,Arco:2020ucn}, and  in \refse{sec:AllowedBR} we discuss the possible allowed values for $\br\left(h\to bs\right)$.
In \refse{sec:Prospects} we discuss the possible experimental reach of present and future colliders in the $\br\left(h\to bs\right)$, as compared to the 2HDM predictions.
Finally,  in \refse{sec:conclusions} we present our conclusions. 



\section{The Two Higgs Doublet Model}
\label{sec:2hdm}
We assume the $\cp$ conserving 2HDM. The scalar potential of this model
can be written as~\cite{Gunion:1989we,Aoki:2009ha,Branco:2011iw}:
\begin{equation}
\begin{split}
V_{\rm 2HDM} = m_{11}^2 (\Phi_1^\dagger\Phi_1) + m_{22}^2 (\Phi_2^\dagger\Phi_2) - \msq (\Phi_1^\dagger
\Phi_2 + \Phi_2^\dagger\Phi_1) + \frac{\la_1}{2} (\Phi_1^\dagger \Phi_1)^2 +
\frac{\la_2}{2} (\Phi_2^\dagger \Phi_2)^2 \\
 + \la_3
(\Phi_1^\dagger \Phi_1) (\Phi_2^\dagger \Phi_2) + \la_4
(\Phi_1^\dagger \Phi_2) (\Phi_2^\dagger \Phi_1) + \frac{\la_5}{2}
[(\Phi_1^\dagger \Phi_2)^2 +(\Phi_2^\dagger \Phi_1)^2] \;,
\end{split}
\label{eq:scalarpot}
\end{equation}
where $\Phi_1$ and $\Phi_2$ denote the two $SU(2)_L$ doublets.
To avoid the occurrence of FCNCs
a $Z_2$ symmetry \cite{Glashow:1976nt,Aoki:2009ha} is imposed on the 
scalar potential of the model under which the 
scalar fields transform as:
\begin{align}
  \Phi_1 \to \Phi_1\;, \quad \Phi_2 \to - \Phi_2\;.
  \label{eq:2HDMZ2}
\end{align}
This $Z_2$, however, is softly broken by the $\msq$ term in
the potential in \refeq{eq:scalarpot}. The extension of the $Z_2$ symmetry to the Yukawa
sector forbids
tree-level FCNCs. 
This results in four variants of 2HDM, 
depending on the $Z_2$ parities of the 
fermion fields.
\refta{tab:types} shows the possible couplings between the 
Higgs doublets and the fermions in the four different 2HDM types.

\begin{table}[htb!]
\begin{center}
\begin{tabular}{lccc} 
\hline
  & $u$-type & $d$-type & leptons \\
\hline
type~I & $\Phi_2$ & $\Phi_2$ & $\Phi_2$ \\
type~II & $\Phi_2$ & $\Phi_1$ & $\Phi_1$ \\
type~III (X or lepton-specific) & $\Phi_2$ & $\Phi_2$ & $\Phi_1$ \\
type~IV (Y or flipped) & $\Phi_2$ & $\Phi_1$ & $\Phi_2$ \\
\hline
\end{tabular}
\caption{Allowed fermion couplings in 
the four types of 2HDM.}
\label{tab:types}
\end{center}
\end{table}

We will study the 2HDM in the ``physical basis",  where the input parameters
of the model are chosen to be:
\begin{equation}
c_{\be-\al} \; , \quad \tb \;, \quad v \; ,
\quad \Mh\;, \quad \MH \;, \quad \MA \;, \quad \MHp \;, \quad m_{12}\equiv\sqrt{\msq} \;.
\label{eq:inputs}
\end{equation}
Here $\tb := v_2/v_1$ is the ratio of the two vacuum expectations values (vevs).
  $\al$ is the mixing angle diagonalizing the $\cp$-even Higgs sector.
  $v := \sqrt{v_1^2 + v_2^2} \approx 246 \gev$ is the SM vev.
From now on we use sometimes the short-hand notation $s_x = \sin(x)$, $c_x = \cos(x)$.
In our analysis we will identify the lightest $\cp$-even Higgs boson,
$h$, with the one observed at $\sim 125 \gev$. 
Also in some occasions we will use $\bar m$,  a derived parameter from $m_{12}$ and $\tb$,  defined as,
\begin{equation}	
\bar m^2=\frac{m_{12}^2}{\sin\beta\cos\beta}\; .
\label{eq:mbar}
\end{equation} 

In the 2HDM, the couplings of the neutral Higgs bosons to SM particles are modified
w.r.t.\ the SM Higgs-coupling predictions due to the mixing in the Higgs
sector.  
Therefore, it is convenient to express the couplings of the neutral scalar
mass eigenstates $h_i$ ($= h,\, H,\, A$) normalized to the corresponding SM couplings.
We introduce the coupling coefficients $\xi_V^{h_i}$ such that the couplings to the massive vector bosons
are given by:
\begin{equation}
\left(g_{h_i W W}\right)_{\mu\nu} =
i g_{\mu\nu}\ \xi_V^{h_i}\ g m_W
\quad \text{and } \quad
\left(g_{h_i Z Z}\right)_{\mu\nu} =
i g_{\mu\nu}\ \xi_V^{h_i}\ \frac{g m_Z}{c_W} \, ,
\end{equation}
where $g$ is the $SU(2)_L$ gauge coupling, $c_W$ is the cosine of weak
mixing angle, $c_W = \MW/\MZ, s_W = \sqrt{1 - c_W^2}$,
and $\MW$ and $\MZ$ denote the masses of the $W$ boson
and the $Z$ boson, respectively. 
For the $\cp$-even boson couplings one finds that $\xi_V^{h}=\SBA$ and $\xi_V^{H}=\CBA$, whereas the coupling to the $\cp$-odd boson is $\xi_V^{A}=0$.

In the Yukawa sector the discrete $Z_2$ symmetry leads to the following
Lagrangian: 
\begin{eqnarray}
	-\mathcal{L}_\mathrm{Yuk} &=&\sum_{f=u,d,l}\frac{m_f}{v}\left[\xi^h_f\bar{f}fh + \xi^H_f\bar{f}fH +\xi^A_f\bar{f}\gamma_5fA \right] \nonumber \\
	& +&\frac{\sqrt{2}}{v} \left[\bar{u}\left(\xi_d V_{\mathrm{CKM}} m_d P_R - \xi_u m_u V_{\mathrm{CKM}} P_L  \right) d H^+ + \xi_l\bar{\nu}m_lP_{R}lH^{+}+\mathrm{h.c.}\right], 
\label{eq:Lag-xi}
\end{eqnarray}
where the parameters $\xi_f^{h,H,A}$, with $f=u,\,d,\,l$, are defined as:
\begin{eqnarray}
&\xi_f^h = s_{\beta-\alpha}+\xi_f c_{\beta-\alpha}\,, \\
&\xi_f^H = c_{\beta-\alpha}-\xi_f s_{\beta-\alpha}\,,  \\
&\xi_u^A = -i \xi_u\;\; \mathrm{and}\;\; \xi_{d,l}^A=i\xi_{d,l}\,.
\end{eqnarray}
The particular values of $\xi_{f}$ for the four 2HDM types shown in \refta{tab:xi}.
Notice that the couplings $\xi_{f}^{h,H}$ of the $\cp$-even Higgs boson coupling to fermions can be understood as the value of the coupling normalized to SM prediction.
\begin{table}[t]
\begin{center}
\begin{tabular}{c|c|c|c|c}
 & Type I & Type II & Type III/Flipped/Y & Type IV/Lepton-specific/X\tabularnewline
\hline 
$\xi_{u}$ & $\cot\beta$ & $\cot\beta$ & $\cot\beta$ & $\cot\beta$\tabularnewline
\hline 
$\xi_{d}$ & $\cot\beta$ & $-\tan\beta$ & $-\tan\beta$ & $\cot\beta$\tabularnewline
\hline 
$\xi_{l}$ & $\cot\beta$ & $-\tan\beta$ & $\cot\beta$ & $-\tan\beta$\tabularnewline
\end{tabular}
\caption{Values for $\xi_{f}$ in the four $Z_{2}$ conserving 2HDM types. }
\label{tab:xi}
\end{center}
\end{table}

The potential in \refeq{eq:scalarpot} defines the interactions in the scalar sector, also involving the SM-like Higgs boson.
We define the triple Higgs couplings $\la_{h h_i h_j}$ couplings such that the Feynman rules are given by $- i\, v\, n!\; \la_{h h_i h_j}$, where $n$ is the number of identical particles in the vertex.
Concretely, the coupling $hH^+H^-$, playing a dominant role in this work, is given by:
\begin{equation}
  \lambda_{hH^+H^-} = \frac{1}{v^2} \Big[\left(m_h^2+2 m_{H^\pm }^2-2
                     \bar{m}^2\right)s_{\be -\al }  
	+ 2 \cot 2 \be \left(m_h^2-\bar{m}^2\right) c_{\be -\al }\Big]\,.
\label{eq:lambda}
\end{equation}

In the so-called alignment limit, where $\CBA\to0$, all $h$ interactions present in the SM tend to their SM values at tree level.  However, even in the alignment limit some BSM interactions can be non-zero, such as $hH^+H^-$ or $HZA$.


\section{\boldmath{$\Gamma\left(h\to bs\right)$} in the 2HDM}
\label{sec:Gamma}

In this section we will analyze the partial decay width of $h\to bs$ in the 2HDM with a softly broken $Z_2$ symmetry. 
Within this model, this observable was already computed for the first time in \citere{Arhrib:2004xu}, prior to the Higgs-boson discovery.
In this paper we update that prediction with the knowledge that there is a SM-like Higgs boson with a mass $\sim 125\gev$.
Furthermore,  we will pay special attention to the effects of the triple Higgs coupling $\lahHpHm$ on the prediction of $\Gamma(h\to bs)$.

In our model set-up
the only non-conserving $\CP$ effect in this process will come from the $\CP$ violating phase in the CKM matrix, that it is known to be very small. 
Therefore, we will compute solely the decay width of the process $h\to b \bar{s}$ and multiply it by 2 to obtain the complete prediction for $\Gamma(h\to bs)$,  to be understood as the sum of 
$\Gamma(h\to b \bar{s})$ and $\Gamma(h\to \bar{b} s)$.
The analytical and numerical results shown in this work were obtained with the help of the public codes {\tt FeynArts} \cite{Hahn:2000kx}, {\tt FormCalc} and {\tt LoopTools}  \cite{Hahn:1998yk}, where the implementation of the 2HDM was made with the public code {\tt FeynRules} \cite{Alloul:2013bka}.
In Appendix~\ref{app:num} we provide more details about the numerical computation of this decay width.

\begin{figure}[t!]
\centering
\includegraphics[width=\columnwidth]{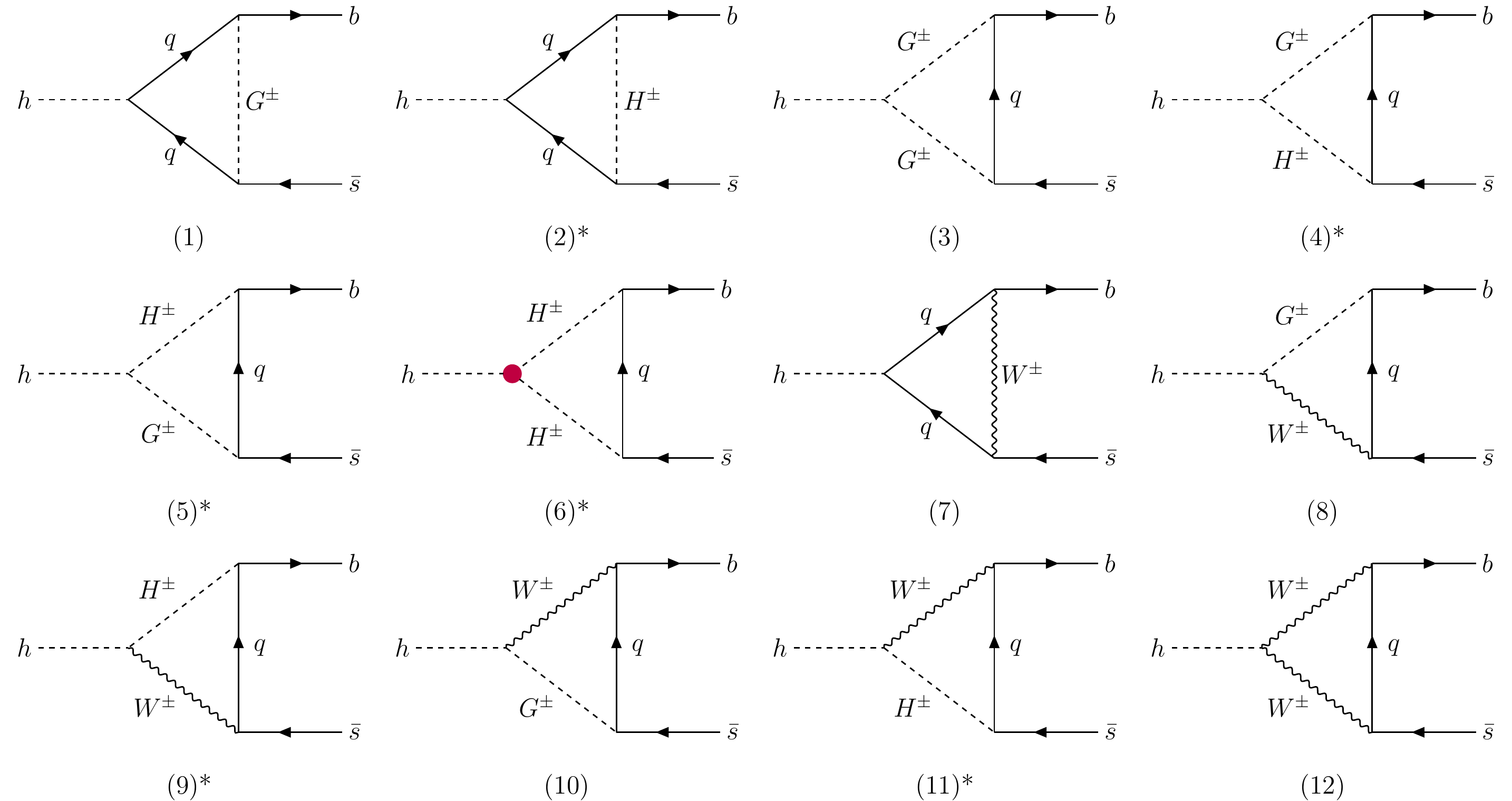}
\includegraphics[width=0.65\columnwidth]{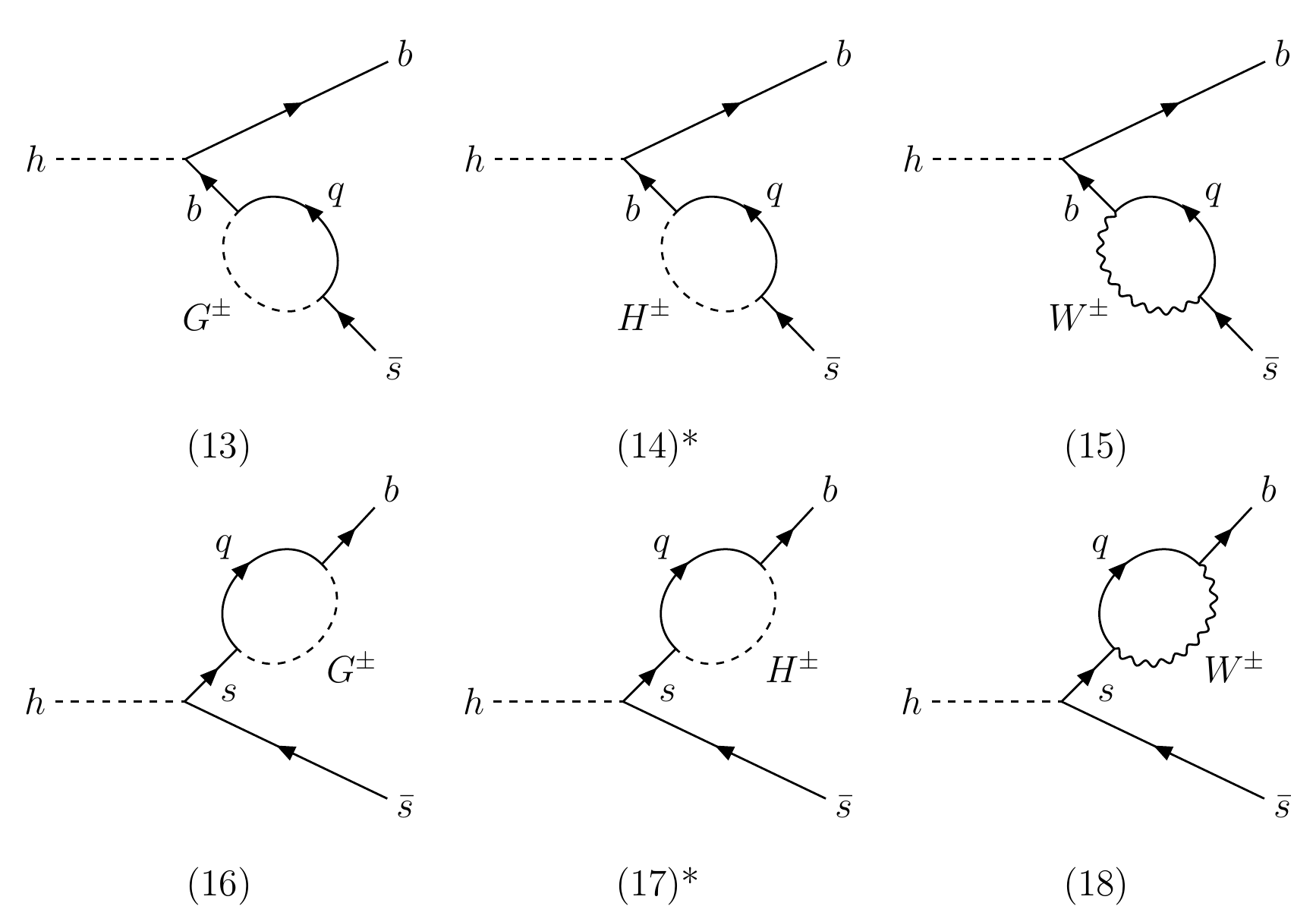}
\caption{One-loop diagrams that contribute to the process $h\to b\bar s$.  
The quark $q$ inside the loops can be any $u$-type quark.  
The diagrams with an asterisk (*) do not appear in the SM.
The red dot marks the $\lahHpHm$ triple Higgs coupling.}
\label{fig:diagrams}
\end{figure}

All one-loop diagrams\footnote{These diagrams were drawn with the help of the public \LaTeX\ code {\tt TikZ-Feynman} \cite{Ellis:2016jkw}.} contributing to the process $h\to b\bar s$ in the 2HDM are depicted in \reffi{fig:diagrams}.
It should be noted that this process is loop induced because the $Z_2$ symmetry imposed in the 2HDM forbids flavor changing neutral currents at tree-level in the scalar sector. 
The total amplitude of the process can be written as:
\begin{equation}
\mathcal{M}\left(h\to b\bar s\right)=\frac{g^3}{64\pi^2m_W^3}\sum_{q=u,c,t}V_{qb}^{\ast}V_{qs}\bar{b}(p_{1})\left(\sum_{i=1}^{18}\mathcal{A}_{i}\right)s(p_{2}),
\label{eq:amp}
\end{equation}
where $V_{ij}$ are
the CKM matrix elements and $\mathcal A_i$ (with $i=1...18$) denotes the contribution to the amplitude of each of the 18 diagrams with the same numbering as shown in \reffi{fig:diagrams}. 
$p_1$ and $p_2$ denote the momenta of the $b$ and $\bar s$ quarks, respectively.
The explicit expressions of the 2HDM Feynman rules involved in the computation and the amplitudes $\mathcal A_i$ in the 2HDM can be found in Appendix~\ref{app:amplitudes}.

It should be noted that the particular fermions that are involved in this process are the various quarks, but not leptons, running in the loops and 
therefore, the prediction of the amplitude in the 2HDM types~I and~IV will be identical.
Likewise,  the prediction in types~II and~III will be also identical (see \refta{tab:xi}).  
The quark $q$ running inside the loops in \reffi{fig:diagrams} can be any $u$-type quark,
accounting for a total of $18\times3=54$ diagrams.
All these diagrams are included in the computation of this decay.  However, numerically some diagrams are more important than others.  
First, all contributions to the amplitude are proportional to either $m_sP_R$ or $m_bP_L$, where $P_{R,L}=(1\pm\gamma^5)/2$ are the right/left chiral projectors. Since $m_s\ll m_b$, the contribution proportional to $P_L$ will dominate.  
On the other hand, concerning the $u$-type quark running inside the loop, the contribution from the $t$ quark is the leading one because it has the larger Yukawa couplings to Higgs bosons and the CKM suppression is milder with respect to other diagrams since $V_{tb}\simeq1$.  
As said above, the computation reported  here is a complete one with the inclusion of all the 54 diagrams present in the 2HDM. 
However,  for illustrative purposes, approximate and simpler expressions considering only the leading contributions to ${\cal A}_i$ described above can be also found in Appendix~\ref{app:amplitudes}.

Finally, another important contribution in the 2HDM to the $h\to b s$ decay is the one coming from
diagram 6, that depends on the triple Higgs coupling $\lahHpHm$.  In our convention, this triple Higgs coupling is dimensionless,  such that the Feynman rule of the vertex $hH^+H^-$ is given by $-iv\lahHpHm$. 
This coupling is a derived parameter within our setup of the 2HDM and its expression as a function of the input parameters $\Mh$, $\MHp$,  $m_{12}$, $\tb$ and $\CBA$ can be found in \refeq{eq:lambda}.
It should be noticed that the coupling $\lahHpHm$ does not vanish in the alignment limit,  i.e. setting $\CBA=0$, and therefore neither does diagram~6.
We will see later that this diagram and its interference effects
with the rest of the diagrams is of crucial importance to the prediction of the decay width of $h\to bs$.

One key fact of the process $h\to b s$ is that it is very suppressed due to the unitarity of the CKM matrix.  
In particular, the unitarity relation that suppresses this decay is:
\begin{equation}
\sum_{q=u,c,t} V_{qb}^\ast V_{qs} = 0\,.
\label{eq:unitCKM}
\end{equation} 
Therefore,  any contribution in the amplitudes ${\cal A}_i$ that does not depend on the mass of the $u$-type quark inside the loop, 
will factor out from the amplitude and vanish because of the unitarity of the CKM matrix.
In order to find these $m_q$-independent parts of the amplitude,  is crucial to perform the Passarino-Veltman reduction to scalar one-loop integrals in the amplitude. 
Thus, these terms in the amplitude were removed explicitly from the numerical evaluation.

Since this is a loop induced process, the full one-loop
result should be free of ultraviolet divergences. 
The only non-divergent diagrams
are diagrams~3 to~7,  both included,  and diagram 12, whereas the remaining diagrams are UV~divergent. 
The total divergence in the amplitude sums up to: 
\begin{equation}
\begin{split}
\mathrm{div}(\mathcal{M})=\frac{g^3}{64\pi^{2}m_{W}^{3}}\sum_{q=u,c,t}V_{qb}^{\ast}V_{qs} \bar b \left(p_1\right) \left(m_{b}P_{L}+m_{s}P_{R}\right) s\left(p_2\right) \\ 
\times
\Delta_\mathrm{UV}\left[m_{W}^{2}s_{\beta-\alpha} 
- \left(m_{q}^{2}\left(\xi_{d}-\xi_{u}\right)\left(\xi_{d}\xi_{u}+1\right)-\xi_{d}m_{W}^{2}\right)c_{\beta-\alpha}\right],
\end{split}
\label{eq:div}
\end{equation}
where $\Delta_\mathrm{UV}=1/\epsilon=2/(4-D)$, $D=4-2\epsilon$ is the number of dimensions, and $\epsilon$ is a small parameter close to zero.
If $\CBA=0$, one recovers the same divergence as in the SM, which vanishes thanks to the unitarity of the CKM matrix, in particular due to \refeq{eq:unitCKM}.
Similarly, in the generic case with $\CBA \neq 0$
the terms proportional to $m_W^2$ also vanish because of the unitarity properties of the CKM matrix. The remaining terms proportional to $m_q^2$ vanish because the relation $\left(\xi_{d}-\xi_{u}\right)\left(\xi_{d}\xi_{u}+1\right)=0$ always holds in a $Z_2$ conserving 2HDM, see for instance \refta{tab:xi}. 

\medskip
The partial width of the process $h\to bs$ can be finally obtained as:
\begin{equation}
\Gamma(h\to bs) \equiv \Gamma(h\to b\bar{s})+\Gamma(h\to \bar{b}s)\simeq 2\times\Gamma(h\to b\bar{s})=2 N_{C}\frac{\lambda^{1/2} \left(m_h^2,m_b^2,m_s^2\right)}{16\pi m_{h}^3}\left|\mathcal{\bar{M}}\right|^{2},
\label{eq:TotalWidth}
\end{equation}
where $N_{C}=3$ is a color factor, 
$\lambda\left(x,y,z\right)=\left(x-y-z\right)^2-4y^2z^2$
 and $\left|\mathcal{\bar{M}}\right|^{2}=\sum_{\mathrm{spin}}\left|\mathcal{M}\right|^{2}$ is the spin-averaged squared amplitude of \refeq{eq:amp}. 
It should be noted that we have neglected the possible $\cp$ violating effects in this process, as we discussed at the beginning of this section.


\subsection{\boldmath{$\Gamma\left(h\to bs\right)$} in the alignment limit}
\label{sec:Gamma-align}

In this subsection,  for illustrative purposes and to simplify the analysis, we will first restrict our study to the alignment limit, i.e. $\CBA=0$.  Under this condition,  the tree-level interactions of $h$ with fermions, gauge bosons and
Goldstone bosons recover their SM value.  
Therefore, diagrams containing SM particles and
the SM-like Higgs boson $h$ (i.e.  diagrams without asterisk in \reffi{fig:diagrams}) are as in the SM.  
In consequence,  in the alignment limit,  the new contributions to $h\to bs$ within the 2HDM arise only from  diagrams with $H^\pm$ propagating in the loops. 
Hence, we can write the amplitude of $h\to bs$ as:
\begin{equation}
\mathcal{M}_
{\rm align}\left(h\to b\bar s\right)=\frac{g^3}{64\pi^2m_W^3}\sum_{q=u,c,t}V_{qb}^{\ast}V_{qs}\bar{b}(p_{1}) \bigg[ {\cal A}_{\rm SM} + {\cal A}_{H^\pm}  \bigg]   s(p_{2}),
\label{eq:ampalign}
\end{equation}
where  ${\cal A}_{\rm SM}$ is the contribution from the SM-like diagrams in the alignment limit and ${\cal A}_{H^\pm}$ is the contribution from diagrams 2, 6, 14 and 17, those that feature a $H^\pm$ propagator in the loop.
The remaining diagrams 4, 5, 9 and 11 vanish in the alignment limit.

In addition,  for comparison,  we have also computed the SM decay width under the same considerations described in the previous section and we have obtained the following value:
\begin{equation}
\Gamma_{\rm SM}\left( h\to bs \right)= 4.22 \times 10^{-10} \gev,
\label{eq:GaSM}
\end{equation}
which is in agreement with the predictions in \citeres{Benitez-Guzman:2015ana,Aranda:2020tqw}.  We will use this value throughout this work to compare consistently the 2HDM prediction with the SM. 
 
One interesting question here is, for which combination of 2HDM input parameters the SM prediction is recovered, corresponding to ${\cal A}_{H^\pm}=0$, see \refeq{eq:ampalign}.
One could expect naively that this ${\cal A}_{H^\pm}$ contribution should vanish for a very heavy $H^\pm$,  since it appears suppressed by a factor $\MHp^{-2}$ from each $H^\pm$ propagator. 
In fact, that is the case for the contributions from diagrams 2, 14 and 17.  
However, diagram 6 could exhibit some non-decoupling effects given the fact that $\lahHpHm$ contains a term proportional to $\MHp^2$. This coupling is non-zero in the alignment limit  and has the following expression (see \refeq{eq:lambda}):
\begin{equation}
  \lambda_{hH^+H^-}^\mathrm{align} = \frac{m_h^2+2 m_{H^\pm }^2-2
                     \bar{m}^2}{v^2} .
\label{eq:alignla}
\end{equation}
For a very large value of $\MHp$ the triangle loop function in diagram~6 scales $\propto \MHp^{-2}$ (see \refeqs{eq:A6}, (\ref{C0-diag6}),  (\ref{C1-diag6}) and  (\ref{C2-diag6})),  whereas the term in $\lahHpHm\propto\MHp^2$  gives rise to a  non-vanishing contribution,  which is constant with $\MHp$ in this heavy mass limit%
 \footnote{It should be noted that this leads to non-perturbative values of $\lahHpHm$ for large values of $\MHp$.}.
To suppress this contribution, one should then tune the prediction of $\lahHpHm$ via a choice for $m_{12}$ such that $\lahHpHm$ does not contain a term that depends quadratically on $\MHp$.  
Nonetheless, it should be noted that outside the alignment limit one can still find BSM effects even with a heavy $H^\pm$ since the interaction $hH^+G^-$ is also proportional to $\MHp^2$. 
This coupling can only be zero if $\CBA=0$ and/or if $\MHp=m_h\simeq 125\gev$.

\begin{figure}[t!]
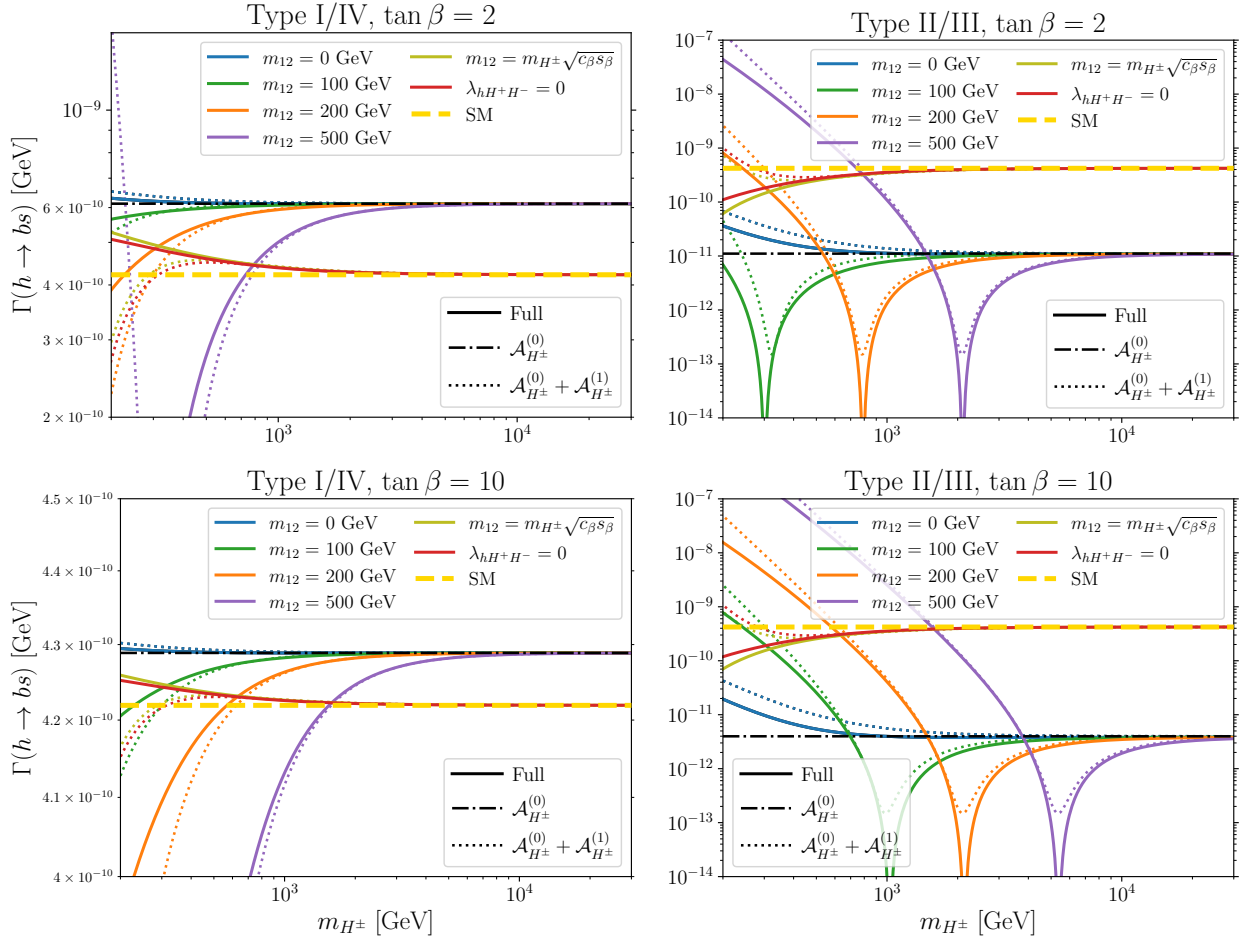

\centering
\includegraphics[width=0.515\textwidth]{plot-yuk1tb2-cbma0_0-A}
\includegraphics[width=0.472\textwidth]{plot-yuk2tb2-cbma0_0-A}
\includegraphics[width=0.515\textwidth]{plot-yuk1tb10-cbma0_0-A}
\includegraphics[width=0.472\textwidth]{plot-yuk2tb10-cbma0_0-A}
\caption{
Prediction of $\Gamma(h\to bs)$ in the 2HDM types I and IV (left) and types II and III (right) as a function of $\MHp$ in the alignment limit for $\tb=2$ (top) and $\tb=10$ (bottom) and for different values of $m_{12}$.  
Dot-dashed and dotted lines correspond to the 2HDM prediction of $\Gamma(h\to bs)$ with the charged Higgs boson effects  approximated by a large $m_{H^\pm}$ expansion given by an effective vertex 
$V^{\rm eff}_{H^{\pm}}$, \refeq{eq:effvertex} with ${\cal
    A}_{H^\pm}^{\left(0\right)}$ (for the choices of fixed $m_{12}$) and ${\cal A}_{H^\pm}^{\left(0\right)}+{\cal A}_{H^\pm}^{\left(1\right)}$ respectively. Dashed yellow horizontal lines indicate the SM prediction.}
\label{fig:dec}
\end{figure}

The numerical prediction of $\Gamma(h\to bs)$ as a function of $\MHp$, in the alignment limit,  is shown in \reffi{fig:dec} for $\tb=2$ (top) and $\tb=10$ (bottom) for 2HDM types I/IV (left) and types II/III (right).
The different solid color lines indicate several choices of the $m_{12}$ parameter.
The SM prediction from \refeq{eq:GaSM} corresponds to the yellow dashed horizontal line. 
In this figure, one can see that the lines corresponding to $m_{12}=0,\,100,\,200,\,500\gev$ tend at large $\MHp$ to a fixed value  of $\Gamma(h\to bs)\neq\Gamma_\mathrm{SM}(h\to bs)$. 
As we discussed above,  this is because  the non-decoupling contribution from diagram~6 due to the presence of $\lahHpHm$. 
That constant value of $\Gamma\left(h\to bs\right)$ reached for large values of $\MHp$ is larger than the SM value in types I/IV and smaller in types II/III. 
This implies that there is a constructive (destructive) interference between
diagram~6 and the rest of the diagrams in types I/IV (II/III) for very large values of $\MHp$.
Hence,  those negative interference effects lead to the deep dips found in the prediction of the partial width in types II/III.
Conversely,  regions with $\bar{m} \gsim \MHp$ which imply $\lahHpHm\lsim0$
 show 
smaller $\Gamma \left( h\to bs \right)$ in the 2HDM than in the
 SM in types I/IV, while in types II/III the 2HDM rates are larger than in the SM,  probing these interference patterns.
These different patterns among the various types can traced back to the different signs of $\xi_d$ between 2HDM types I/IV and II/III, that enter in the Yukawa interactions of $H^\pm$ (see \refeq{eq:Lag-xi} and \refta{tab:xi}).

On the other hand,  the red solid line in each plot of \reffi{fig:dec} corresponds to a value of $m_{12}$ such that the coupling $\lambda_{hH^+H^-}$ in \refeq{eq:alignla} is equal to zero%
\footnote{In the alignment limit this corresponds to $m_{12}^2=\sin\beta\cos\beta\left(\MHp^2+m_h^2/2 \right)$ or $\bar m^2=\MHp^2+m_h^2/2$.}. 
Therefore in that case the only remnant 2HDM contribution in $\mathcal{A}_{H^\pm}$ comes from diagrams 2, 14 and 17.
For all 2HDM types,  this line reach the SM prediction  for $\MHp\gtrsim2\tev$, just as expected.
The olive line indicates the prediction for the choice $m_{12}=\MHp\sqrt{\cos\beta\sin\beta}$ (or equivalently $\bar m = \MHp $),  which implies the particular setting of $\lahHpHm=\Mh^2/v^2$.
In this case, since $\lahHpHm$ does not depend on $\MHp$,  there is not a non-decoupling effect from diagram 6 and the prediction for $\Gamma\left(h\to bs\right)$ also reaches the SM value for sufficiently large $\MHp$. 

In fact, the appearing gap at large  $\MHp $ between the predictions with a fixed value of $m_{12}$ and the red line with $\lahHpHm=0$, or the olive line with $\lahHpHm = \Mh^2/v^2$ highlights the  non-decoupling contribution from diagram~6 for parameter choices that lead to $\lahHpHm \propto \MHp^2$.
It also confirms that the interference effects discussed above are caused by the presence of diagram~6, and consequently, the value of $\lahHpHm$ is crucial in the 2HDM prediction of  $\Gamma\left(h\to bs\right)$.


\subsection{One-loop effective \boldmath{$hbs$} vertex from the large \boldmath{$\MHp$} expansion}
\label{sec:largeMHp}

Given the presence of possible decoupling and non-decoupling effects in $\Gamma\left(h\to bs\right)$ described in the previous section, here  
we present a large $\MHp$ series expansion that will capture such effects.  
We work again in the alignment limit, i.e,  we set $\CBA=0$.  The analytical results will be presented in the form of a one-loop effective vertex,  $V^{\rm eff}_{H^\pm}$,  for the $h b s$ interaction generated from integrating out the heavy mode $H^\pm$,  which  is valid at large 
$\MHp $. Such an effective vertex can be useful to obtain a rough estimate of the most relevant numerical contributions from $H^\pm$  to the $h \to b s$ decay rates in the 2HDM.

In order to obtain this effective vertex, we have performed a series expansion of the 2HDM contribution $\mathcal{A}_{H^\pm}^{\rm eff}$ in powers of a small dimensionless variable $x_{H^\pm}$,  assuming that 
\begin{equation}
x_{H^\pm}=m_{\rm EW}^2/\MHp^2\ll1,
\end{equation}
 where $m_{\rm EW}$ denotes generically any mass-dimension parameter (other than $\MHp$)  involved in the computation, i.e.  $m_h$, $m_W$, $m_q$ and $m_{12}$.
Furthermore, for the results in this section,  we have considered only the dominant contributions to ${\cal A}_{H^\pm}$,  which are globally proportional to $m_t^2 m_b$. 
In summary,  the validity of our assumptions in this section applies in
  practice to the following mass hierarchy: $\MHp \gg m_t, \ m_h,\ m_{12} \gg
  m_b \gg m_s,\ m_c,\ m_u$. (The case $\MHp \sim m_{12}$ will be
    discussed separately below.) This hierarchy, on the other hand, 
 (as discussed in the beginning of \refse{sec:Gamma}) implies that the result for the effective vertex  will only involve the left projector $P_L$,  since the $P_R$ contribution is always proportional to $m_s$ and can therefore be safely neglected.  The analytical computation presented below
was performed with the help of the public code {\tt PackageX} \cite{Patel:2015tea}.

We have computed the first two contributions in this series expansion, 
\begin{equation}
\mathcal{A}_{H^{\pm}}^{\rm eff}=\mathcal{A}_{H^{\pm}}^{\left(0\right)}+\mathcal{A}_{H^{\pm}}^{\left(1\right)}+ \dots
\label{eq:expan}
\end{equation}
where $\mathcal{A}_{H^{\pm}}^{\left(0\right)}$ is the leading order contribution,  i.e.  $\mathcal{O}(x_{H^\pm}^0)$, and $\mathcal{A}_{H^{\pm}}^{\left(1\right)}$ is the next to leading order contribution, i.e.  $\mathcal{O}(x_{H^\pm}^1)$.  The dots in the above equation correspond to terms of $\mathcal{O}(x_{H^\pm}^n)$, with $n=2,3,..$.  
In the following of this subsection we first present the analytical results of $ \mathcal{A}_{H^{\pm}}^{\left(0\right)}$ and $ \mathcal{A}_{H^{\pm}}^{\left(1\right)}$ in the generic case where 
$ \lambda_{hH^+H^-}$ has been replaced by the expression in the alignment limit given by \refeq{eq:alignla}.
Subsequently, we compare them with the analytical results of the two cases with the particular choices of $ \lambda_{hH^+H^-}=0$ and $ \lambda_{hH^+H^-}=m_h^2/v^2$.

The result obtained for the leading order contribution is the following:
\begin{equation}
\mathcal{A}_{H^{\pm}}^{\left(0\right)}=m_{t}^{2}m_{b}P_{L}\xi_{u}\biggl\{\frac{1}{2}\xi_{u}-2\xi_{d}\biggr\}\,.
\label{eq:order0}
\end{equation}
This is the non-decoupling contribution (valid under the above given assumptions) discussed in the previous section that comes solely from diagram~6.
With the particular values of $\xi_u$ and $\xi_d$ from \refta{tab:xi} in  \refeq{eq:order0}, we get the following non-decoupling effects for the four 2HDM types:
\begin{equation}
 \mathcal{A}_{H^{\pm}}^{\left(0\right)}=
\begin{cases}
-\frac{3}{2}m_t^2m_bP_L\cot^2\beta & \mathrm{Types\ I/IV},  \vspace{2mm} \\
\frac{1}{2}m_t^2m_bP_L\left(4+\cot^2\beta\right)  & \mathrm{Types\ II/III}.
\end{cases}
\label{eq:dectypes}
\end{equation}
Hence,  $\mathcal{A}_{H^\pm}^{\left(0\right)}$ has different signs in types~I/IV and~II/III in the large $\MHp$ limit. 

The result for the next to leading order contribution is  the following: 
\begin{equation}
\begin{split}
& \mathcal{A}_{H^{\pm}}^{\left(1\right)}=\frac{m_{t}^{2}m_{b}P_{L}\xi_{u}}{36m_{H^{\pm}}^{2}}\left\{ 36\bar{m}^{2}\left(2\xi_{d}-\frac{1}{2}\xi_{u}\right)+m_{h}^{2}(11\xi_{u}-46\xi_{d})-120m_{t}^{2}\xi_{d}\right.\\ & \left.+12\xi_{d}\left(2\left(m_{h}^{2}+5m_{t}^{2}\right)\sqrt{4r-1}\ \mathrm{arccot}\left(\sqrt{4r-1}\right)-m_{h}^{2}\log\left(\frac{m_{H^{\pm}}^{2}}{m_{t}^{2}}\right)\right)\right\}\,,
\end{split}
\label{eq:order1}
\end{equation}
with $r=m_t^2/m_h^2$.  
In the above result, the effects from $H^\pm$  are suppressed by $\MHp^{-2}$ and come from diagrams 2,  6 and 14. 
Diagram 17 is proportional to $m_s$ and therefore negligible in our approximations.
A more detailed derivation of the above equations can be found in Appendix~\ref{app:expansion}.

In summary,  and for practical purposes,  the effects from the  $H^{\pm}$  in the alignment limit to the $h\to bs$ decay rates can be approximated by a one-loop effective $hbs$ vertex $V^{\rm eff}_{H^{\pm}}$, whose value,  according to \refeq{eq:amp},  is given by:
\begin{equation}
\begin{gathered}
\includegraphics[scale=.9]{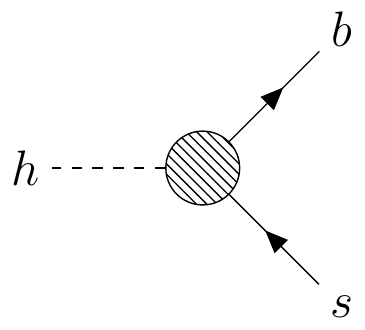}
\end{gathered}
:
\
\
\
\
V^{\rm eff}_{H^{\pm}}
\
=
\
\frac{g^3}{64\pi^2m_W^3}V_{tb}^{\ast}V_{ts}\;{\cal A}_{H^\pm}^{\rm eff}\; ,
\label{eq:effvertex}
\end{equation}
where ${\cal A}_{H^\pm}^{\rm eff}$,  as defined in \refeq{eq:expan},  receives contributions from leading order and next to leading order in $x_{H^\pm}$ as given in \refeqs{eq:order0}-(\ref{eq:order1}). 

The approximate prediction for $\Gamma\left(h\to bs\right)$ with the effects of the heavy $H^\pm$ described by the effective vertex in
  \refeq{eq:effvertex} is also shown in \reffi{fig:dec} for the cases of
    fixed $m_{12}$ values.
Both results  with $V^{\rm eff}_{H^{\pm}}$ given by the leading order
contribution,  ${\cal A}_{H^\pm}^{\left(0\right)}$ (dot-dashed lines),  and by
the sum of the leading and next to leading contributions,  ${\cal
  A}_{H^\pm}^{\left(0\right)}+{\cal A}_{H^\pm}^{\left(1\right)}$ (dotted
lines)  are shown for the different fixed values of $m_{12}$. 

One can see that the approximation given by  ${\cal A}_{H^\pm}^{\left(0\right)}$ already works quite well, specially for large values of $\MHp$,  just as expected.
For example,  for $m_{12}=0$,  this approximation is very close to the full prediction in type I/IV for values of $\MHp$ around hundreds of GeV while for types II/III the approximation works fine from around 1~TeV.
As expected, for the lines with larger values of $m_{12}$, it is required to go to larger values of $\MHp$ in order that ${\cal A}_{H^\pm}^{\left(0\right)}$  be a good approximation of ${\cal A}_{H^\pm}$.
Also,  this leading order approximation can reproduce quite well the dependence of the non-decoupling contributions for different values of $\tb$.
In fact, this leading contribution can already describe that the gap between the SM and the 2HDM non-decoupling value increases with $\tan\beta$ in types II/III and decreases in types I/IV.

Turning to the prediction with the effective vertex including both leading and next to leading order contributions,  ${\cal A}_{H^\pm}^{\left(0\right)}+{\cal A}_{H^\pm}^{\left(1\right)}$,  we also see in \reffi{fig:dec} that it approximates better the full result than considering just the leading order, in particular in the low $\MHp$ region where the ${\cal O}(m_{\rm EW}^2/\MHp^2)$ corrections are more relevant.
Furthermore,  in types II/III, it also reproduces with a good accuracy the location of the deep interference minima present in these types, particularly if it takes place for a large value of $\MHp$.

Finally,  we comment on the effective vertex approximation for choices of $m_{12}$ that do not lead to the
  dependence of $\lahHpHm \propto \MHp^2$ for large values of $\MHp$. In
  \reffi{fig:dec} this is realized for the particular choice of $m_{12}^2=\MHp^2 \sin\beta\cos\beta$ that leads to  $\lahHpHm=m_h^2/v^2$,  and for the choice of $m_{12}^2=\sin\beta\cos\beta\left(\MHp^2+m_h^2/2 \right)$ that leads to $\lahHpHm=0$.
For these parameter choices,  $m_{12}^2$ is constrained to contain a
  term  directly proportional to $\MHp^2$, i.e.,  the required expansion condition $m_{12}\ll\MHp$ shown in \refeq{eq:expan} is not satisfied.
Therefore, the consistent way to perform the series expansion in these cases is to consider the particular expression for $m_{12}$, and the resulting value of $\lahHpHm$, before the series expansion.
As noted above, the absence of a term proportional to $\MHp^2$ in $\lahHpHm$ implies complete decoupling in the prediction of $\Gamma(h\to bs)$ for large values of $\MHp$.
Thus, for both scenarios, after doing the expansion, we find a vanishing leading term:
\begin{equation}
	{\cal A}_{H^\pm}^{\left(0\right)}= 0\,,
\end{equation}
which summarizes the absence of non-decoupling effects in these two cases.
However, there are still non-vanishing contributions to ${\cal A}_{H^\pm}^{\left(1\right)}$, which are suppressed by $\MHp^2$ and therefore summarize the decoupling effects in these two particular cases.
If $\lahHpHm=0$, the result for the next to leading order contribution is:
\begin{equation}
\begin{split}
&\mathcal{A}_{H^{\pm}}^{\left(1\right)} = \frac{m_{t}^{2}m_{b}P_{L}\xi_{u}}{36m_{H^{\pm}}^{2}}\bigg\{ -4 m_{h}^{2}\xi_{d} + 6 m_{t}^{2}\left(3 \xi_{u} - 8 \xi_{d} \right) \\ & \left.  + 12\xi_{d}\left(2\left(m_{h}^{2}+5m_{t}^{2}\right)\sqrt{4r-1}\ \mathrm{arccot}\left(\sqrt{4r-1}\right)-\left( m_{h}^{2} +6m_{t}^2 \right) \log\left(\frac{m_{H^{\pm}}^{2}}{m_{t}^{2}}\right)\right)\right\}\,,
\label{eq:order1-la0}
\end{split}
\end{equation}
which only receives contributions from diagrams 2 and 14.
On the other hand, if $m_{12}=\MHp\sqrt{\sin\beta\cos\beta}$, corresponding to $\lahHpHm=\Mh^2/v^2$, the result for ${\cal A}_{H^\pm}^{\left(1\right)}$ also receives contributions from diagram 6. In this case the result is:
\begin{equation}
\begin{split}
& \mathcal{A}_{H^{\pm}}^{\left(1\right)}  = \frac{m_{t}^{2}m_{b}P_{L}\xi_{u}}{36m_{H^{\pm}}^{2}} \bigg\{ m_{h}^{2}\left(9\xi_{u}-40\xi_{d}\right) + 6m_{t}^{2}\left(3\xi_{u}-8\xi_d \right)  \\ & \left.  + 12\xi_{d}\left(2\left(m_{h}^{2}+5m_{t}^{2}\right)\sqrt{4r-1}\ \mathrm{arccot}\left(\sqrt{4r-1}\right)-\left(m_{h}^{2}+6m_{t}^2\right)\log\left(\frac{m_{H^{\pm}}^{2}}{m_{t}^{2}}\right)\right)\right\}\,.
\label{eq:order1-lamh2v2}
\end{split}
\end{equation}
It should be noted that the two expressions above are different from the previous result of \refeq{eq:order1}, as expected.
In that result, there are additional decoupling contributions of $\mathcal{O}\left(\MHp^{-2}\right)$ coming from diagram 6 that are absent if $m_{12}=\MHp\sqrt{\sin\beta\cos\beta}$ or $\lahHpHm=0$.

The numerical results of the effective vertex considering the explicit expressions of $\mathcal{A}_{H^{\pm}}^{\left(1\right)}$ in the effective vertex $V_{H^\pm}^\mathrm{eff}$ are also shown in \reffi{fig:dec}.
The dotted red line shows the result from the approximation in the case that $\lahHpHm=0$ (\refeq{eq:order1-la0}), while the dotted olive line shows the result of the approximation for $\bar m=\MHp$ (\refeq{eq:order1-lamh2v2}).
Since $\mathcal{A}_{H^\pm}^{\left(0\right)}=0$ in both cases, they can reproduce the expected decoupling behavior given these parameter choices for $m_{12}$.
Additionally, the convergence to the SM prediction is also well reproduced by the effective vertex approximation considering the results for $\mathcal{A}_{H^\pm}^{\left(1\right)}$ for both cases. In particular, the approximation yields a very good agreement with the complete prediction when $\MHp\gtrsim 500\gev$.


\subsection{\boldmath{$\Gamma\left(h\to bs\right)$} outside the alignment limit}
\label{sec:Gamma-noalign}

\begin{figure}[p!]
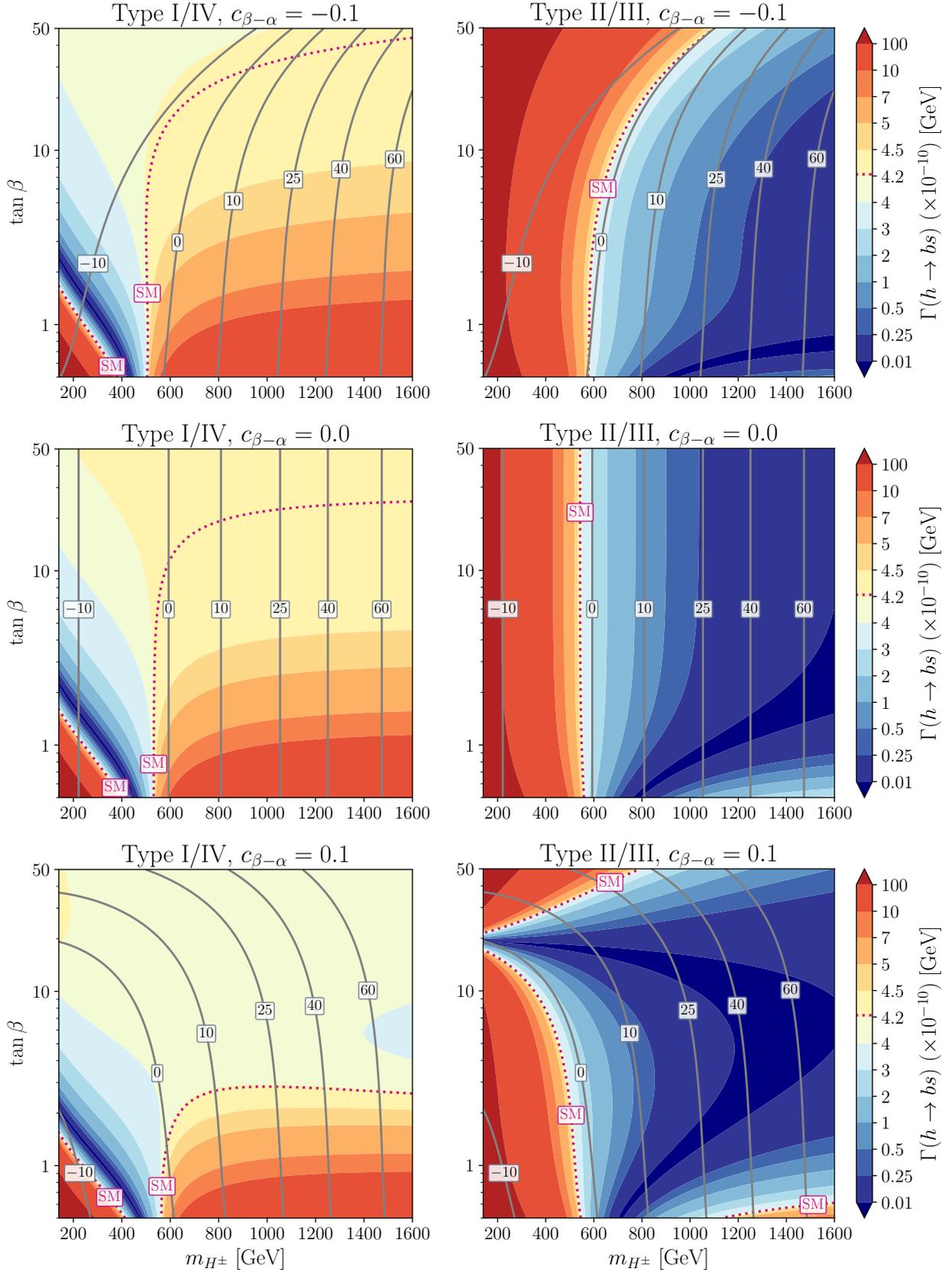

\centering
\includegraphics[width=0.46\textwidth]{plot-yuk1-mbar600-cbma-0_1}
\includegraphics[width=0.525\textwidth]{plot-yuk2-mbar600-cbma-0_1}
\includegraphics[width=0.46\textwidth]{plot-yuk1-mbar600-cbma0_0}
\includegraphics[width=0.525\textwidth]{plot-yuk2-mbar600-cbma0_0}
\includegraphics[width=0.46\textwidth]{plot-yuk1-mbar600-cbma0_1}
\includegraphics[width=0.525\textwidth]{plot-yuk2-mbar600-cbma0_1}
\caption{Prediction of $\Gamma(h\to bs)$ in the plane $\MHp$-$\tb$ with $\bar m=600\gev$ for 2HDM types I/IV (left column) and types II/III (right column). 
$\CBA$ is set to -0.1 (top), 0 (middle) and 0.1 (bottom).
Dotted pink line corresponds to the SM prediction. 
Solid grey lines show contour lines for different values of $\lahHpHm$.}
\label{fig:mHpvstb}
\end{figure}

Once we studied the behavior of $\Gamma(h\to bs)$ within the alignment limit, the next step is to consider this observable outside this limit. The fact that $c_{\beta-\alpha}\neq0$ implies that the couplings of $h$ to the fermions, gauge bosons and Goldstone bosons are not SM-like. Consequently, deviations with respect to the SM can arise from all diagrams. 
The $h$ coupling to fermions scales with $\xi_f^h=\SBA+\xi_f\CBA$ (see \refeq{eq:Lag-xi}).
Regarding the $h$ couplings to gauge and Goldstone bosons,  they are now proportional to $\SBA$ and Yukawa-type independent, and consequently they are always smaller than or equal to the SM Higgs-boson couplings.
 Furthermore, another interactions involving the charged Higgs boson $H^\pm$ are now also present in the prediction, like $hH^\pm G^\mp,\ hH^\pm W^\mp\propto \CBA$, and consequently they will depend on the sign of $\CBA$ and will be potentially small if $\CBA$ is also small.  
Finally,  the last new effect can enter thought $\lahHpHm$, that is a function of $\CBA$ given in \refeq{eq:lambda}.  
Due to these features appearing outside the alignment limit,  it is not possible to split the 2HDM result for the decay amplitude into
 the SM and the $H^\pm$ contributions as in \refeq{eq:ampalign}, and neither to write the (possibly) most relevant contribution from the 2HDM by means of just an effective vertex for very large $\MHp$ as in \refeq{eq:effvertex}.  Thus, we present in this section the numerical results using the full formulas for the decay amplitude and we do not employ anymore the approximate formulas with the effective vertex discussed in the previous section. 
Since we investigate here the analytical dependencies of $\Ga(h \to bs)$ we do not apply any experimental or theoretical constraints. This will be done in the phenomenological analysis in \refse{sec:BR}.

In \reffi{fig:mHpvstb} we show the effects of $\CBA\neq0$ on $\Gamma(h\to bs)$ in the plane $\MHp$-$\tb$ with $\bar m=600\gev$ for the 2HDM types I/IV (left column) and II/III (right column).  
The prediction for $\CBA=-0.1,\;0.0,\;0.1$ is shown for both cases in the top, middle and bottom rows, respectively.
The dotted pink contour lines indicates where the partial decay width has the same value as in the SM.
The grey contour lines in the plots correspond to constant values of the triple Higgs coupling $\lahHpHm$, i.e. the parameter that controls to a large extend the size of the contribution of diagram~6.

We first focus on the plots in the alignment limit as shown in the middle row of \reffi{fig:mHpvstb}. In types I/IV (middle left plot), we see that for $\lahHpHm\lsim 0 \,(\gsim 0)$ very important destructive (constructive) interference effects are found.  
These interference effects are mainly dominated by the effect of diagram 6, the one containing $\lahHpHm$,  just as we found in the previous section.
In fact, there is a narrow region where the prediction of the partial width can be below $10^{-12}\gev$,  just in the region where $\lahHpHm$ becomes negative. 
If we turn now to the prediction in types~II and~III (middle right plot), the situation is the opposite as compared to types~I and~IV. 
Now the strong destructive effects 
are found when $\lahHpHm\gsim0$, while in the region where $\lahHpHm\lsim0$ they are absent. 
It can also be seen that the contour lines become flat for larger values of $\MHp$, these are the non-decoupling effects for fixed $m_{12}$ that,  as discussed in the previous section, can be approximated by \refeq{eq:order0}.
In the case of types II/III it would be necessary to go to larger values of $\MHp$ to reach this flat behavior. 
Furthermore, we can also see the effects of large $\tb$ in these plots. 
In types I/IV, it can be observed that the prediction of $\Gamma(h\to bs)$ exhibits a strong dependence on $\tb$ for lower values of this parameter, whereas the dependence is less pronounced in the region of large $\tan\beta$.
This is because all diagrams with a $H^\pm$ running inside the loop are proportional to $\xi_d=\cot\beta$,  so they are very suppressed in this region of large $\tb$.
Nonetheless, in types II/III, $\xi_d=-\tan\beta$ and these diagrams are relevant when $\tan\beta$ is large. 
In fact, some leading contributions are proportional to $\xi_u\xi_d$, and therefore in these 2HDM types we see $\tan\beta$ independence in the contour lines for $\Gamma\left(h\to bs\right)$.

We turn now to the cases where $\CBA\neq0$. 
We can understand these results as distortions from the alignment limit case.
In types I/IV, we see that overall the partial width is slightly larger (smaller) when $\CBA=-0.1 (+0.1)$ with respect to the alignment limit case,  specially for large values of $\MHp$.
The reason for this behavior is that the SM-like diagrams are slightly enhanced for $\CBA>0$ and diminished for $\CBA>0$, and thus account for these subtle differences.
Now we turn to types II/III, where we see noticeable differences between the plots with $\CBA=\pm0.1$  and the prediction in the alignment limit, contrary to what we have seen in types I/IV.  
As before, the contribution from diagram 6 dominates the pattern of the prediction of $\Gamma\left(h\to bs\right)$ for low and moderate values of $\tb\lsim5$, where the prediction is very similar to the alignment limit case.
However, in the large $\tb$ region there is a new non-negligible 2HDM effect coming from diagram 14,  which can be enhanced by $\xi_d=-\tb$.
In addition, the SM-like diagrams can be also strongly enhanced by the same parameter, like for example diagram 13.
Furthermore, because of the interplay between all these contributions,  the deep minimum due to the negative interference can be found around $\tb\simeq20$ for lower values of $\MHp$ and $\CBA=+0.1$.

It should be noted that the SM prediction, corresponding to the pink dashed contours in \reffi{fig:mHpvstb} is found in both sets of Yukawa types (I and IV vs.\ II and III) in different regions of the parameter space. 
In contrast to our discussion of the alignment limit in \refse{sec:Gamma-align}, in these regions the sum of the SM-like diagrams do not yield the SM value: their non-SM contributions are canceled by the non-SM diagrams involving the $H^\pm$.


\section{Allowed values of \boldmath{$\mathrm{BR}\left(h\to bs\right)$} in the 2HDM }
\label{sec:BR}
After the theoretical study of the prediction of $\Gamma\left(h\to bs\right)$, in this section we will analyze the possible size for the branching ratio still allowed by the current theoretical and experimental constraints of the 2HDM.
The restrictions on the parameter space for the 2HDM was studied recently in~\citere{Arco:2020ucn, Arco:2022xum}. Here we follow the same procedure and only briefly summarize the considered constraints:

\begin{itemize}
\item \textbf{Electroweak precision data.}
New physics effects on electroweak observables from extended Higgs sectors can be parametrized in terms of the oblique parameters  $S$, $T$ and $U$ \cite{Peskin:1990zt,Peskin:1991sw}.
The values for the oblique parameters reported by the PDG \cite{ParticleDataGroup:2022pth} are very close to zero, meaning that the new physics contribution should remain small.
Within the 2HDM, the most constraining oblique parameter is $T$, that is very sensitive to mass splittings between the additional Higgs bosons.  
Therefore,  in the rest of this work, we will consider $\MA=\MHp$,  because this choice keeps the $T$ under control at the one-loop level,  even outside the alignment limit.%
\footnote{We will not consider here possible changes in the determination of the $S$, $T$, $U$ parameters due to the recent measurement of $m_W$ by CDF~\cite{CDF2}.}
\item \textbf{Theoretical constraints.}
  These constraints consists 
in tree-level perturbative unitarity and the stability of the 2HDM potential.
For the tree-level unitarity,  we set an upper bound on the modulus of the eigenvalues of the 
two-body $s$-wave scattering amplitude matrix
between the scalars fields \cite{Kanemura:1993hm,Akeroyd:2000wc,Ginzburg:2005dt}.
For potential stability, we demand the 2HDM potential to be bounded from below \cite{Deshpande:1977rw,Maniatis:2006fs}. 
Additionally,  we also require that the minimum of the potential is a true minimum \cite{Barroso:2013awa}.
All these constraints can be translated to inequalities involving the parameters $\lambda_i$ from the 2HDM potential in \refeq{eq:scalarpot} and can be found in \citere{Arco:2020ucn}.
Typically,  these theoretical constraints impose the hardest bounds on the tripe Higgs couplings size and, in particular, on $\lahHpHm$.  

\item \textbf{Higgs searches at colliders.} 
We require that the Higgs bosons predicted by the 2HDM avoid the known experimental searches for new particles at the 95\% CL.
We made use of the public code {\tt HiggsBounds5.9}~\cite{Bechtle:2008jh,Bechtle:2011sb,Bechtle:2013wla,Bechtle:2015pma,Bechtle:2020pkv,Bahl:2022igd}, that determines if a particular configuration of an extended-Higgs model is excluded at the 95\% CL by the existing experimental bounds.
The theoretical predictions required by {\tt HiggsBounds} to check the experimental bounds were computed with the public code {\tt 2HDMC}~\cite{Eriksson:2009ws}.
{\tt HiggsBounds} contains experimental bounds from LEP,  Tevatron and the LHC, including also all relevant results from the LHC Run~2.

\item \textbf{Rate measurements of the 125 GeV Higgs boson.}
We have identified the $h$ state from the 2HDM with the $~125\gev$ SM-like Higgs boson discovered by the LHC. 
The properties of $h$ should be in agreement with the experimental measurements of the mass and the signal strengths of such Higgs boson.
To evaluate this agreement we made use of the public code {\tt HiggsSignals2.6}~\cite{Bechtle:2013xfa,Bechtle:2014ewa,Bechtle:2020uwn,Bahl:2022igd}, that performs a statistical $\chi^2$ analysis considering all the  Higgs boson 
signal strength and mass
measurements available from Tevatron and the LHC, including most relevant results from the LHC Run~2.
Again, the theoretical predictions needed by {\tt HiggsSignals} to compute the $\chi^2$ fit are obtained with {\tt 2HDMC}.
We will consider the allowed region 
to be not further away than $2\sigma$ ($\Delta\chi^2=6.18$ for a two-dimensional scan, i.e.\ a plane) from the SM, where $\chi^2_{\rm SM}=85.76$ with 107 observables.
It is known that the discovered Higgs boson is in a very good agreement with the SM predictions, therefore the results from {\tt HiggsSignals} will be translated into allowed values of $\CBA$ very close the alignment limit.
Stronger constraints on $\CBA$ are expected in types II, III and IV, where the $h\to bb$ and/or the $h\to\tau\tau$ rates are enhanced by large values of $\tb$ (depending on the type, see \refta{tab:xi}).

\item \textbf{Flavor observables.}
The new Higgs bosons in the 2HDM can induce relevant loop corrections to several rare $b$-flavored meson decays.
In many BSM models the most important new contributions to these $B$-physics observables come from the effects from diagrams mediated by charged Higgs bosons, $H^\pm$.
In our analysis, we will consider the 95\% CL limits from the experimental average of $\br\left(B\to X_s\gamma\right)$ and $\br\left(B_s\to\mu\mu\right)$,
as given in~\cite{ParticleDataGroup:2022pth}.
The theoretical computation of these BRs within the 2HDM was done with the public codes {\tt SuperISO}~\cite{Mahmoudi:2008tp,Mahmoudi:2009zz} and {\tt 2HDMC}.
Stronger bounds are known to be found for all types in the low $\tb$ region with light $H^\pm$. Furthermore, in types II and III, there is a nearly $\tb$-independent bound on $\MHp \gsim 500\gev$.
\end{itemize}


\subsection{Results of \boldmath{$\mathrm{BR}\left(h\to bs\right)$}}
\label{sec:AllowedBR}

The results in this section will be given in terms of the ratio between the $\br\left(h\to bs\right)$ prediction in the 2HDM with respect to the SM,  given by:
\begin{equation}
R_\mathrm{2HDM}=\frac{\br\left(h\to bs\right)_\mathrm{2HDM}}{\br\left(h\to bs\right)_\mathrm{SM}}.
\end{equation}
The computation of the total width of $h$ needed to estimate $\br\left(h\to bs\right)_{\rm 2HDM}$ was done with the public code {\tt 2HDMC} \cite{Eriksson:2009ws}.
For the $\br\left(h\to bs\right)_{\rm SM}$ prediction we used the following value of the total width:
\begin{equation}
\Gamma_\mathrm{SM}\left(h\right)=4.353\times10^{-3} \gev,
\end{equation}
obtained with {\tt 2HDMC} via setting $\CBA=0$ and $\lahHpHm=0$ via a tuned value of $m_{12}$\footnote{The {\tt 2HDMC} code only includes $h$ decays to $q\bar q$ (with QCD corrections),  $l^+l^-$,  $WW^\ast$,  $ZZ^\ast$,  $\gamma\gamma$ and $Z\gamma$, without EW corrections. 
With $\CBA=0$, the $h$ tree-level couplings to SM particles go to their SM values.
With $\lahHpHm=0$,  the new $H^\pm$-mediated diagrams in the $\gamma\gamma$ and $Z\gamma$ decays vanish.  See \cite{Arco:2022jrt} for further details.}.
This choice allows for a consistent comparison of both quantities at the same accuracy level.

\begin{figure}[t!]
\centering
\includegraphics[width=0.4575\textwidth]{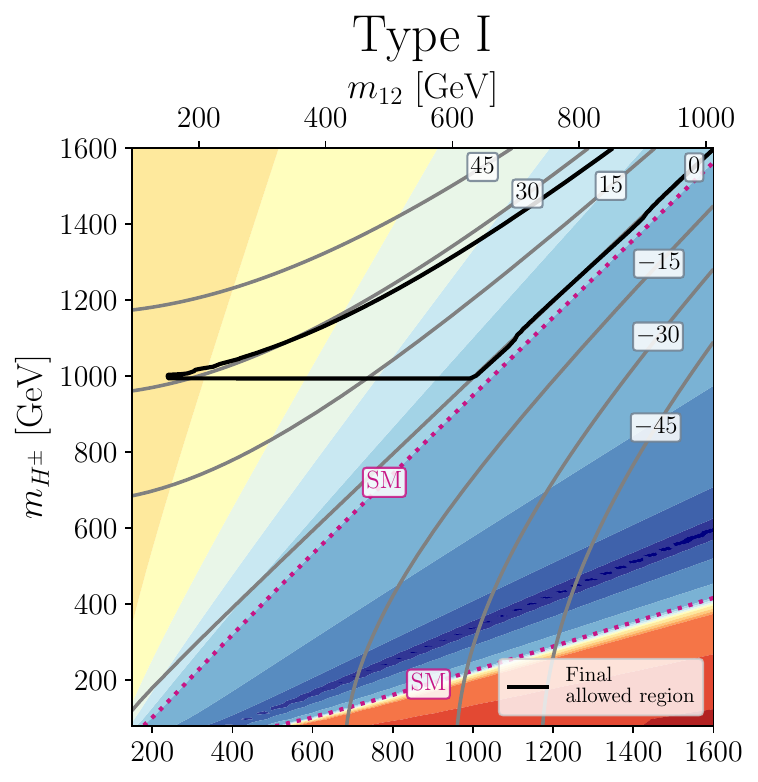}\includegraphics[width=0.54\textwidth]{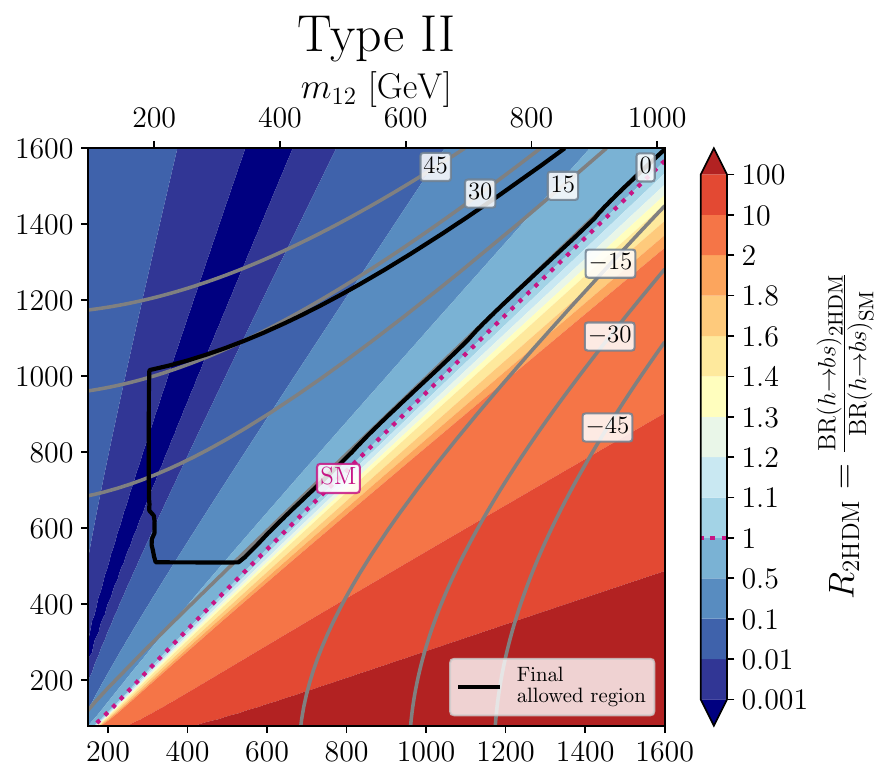}
\includegraphics[width=0.4575\textwidth]{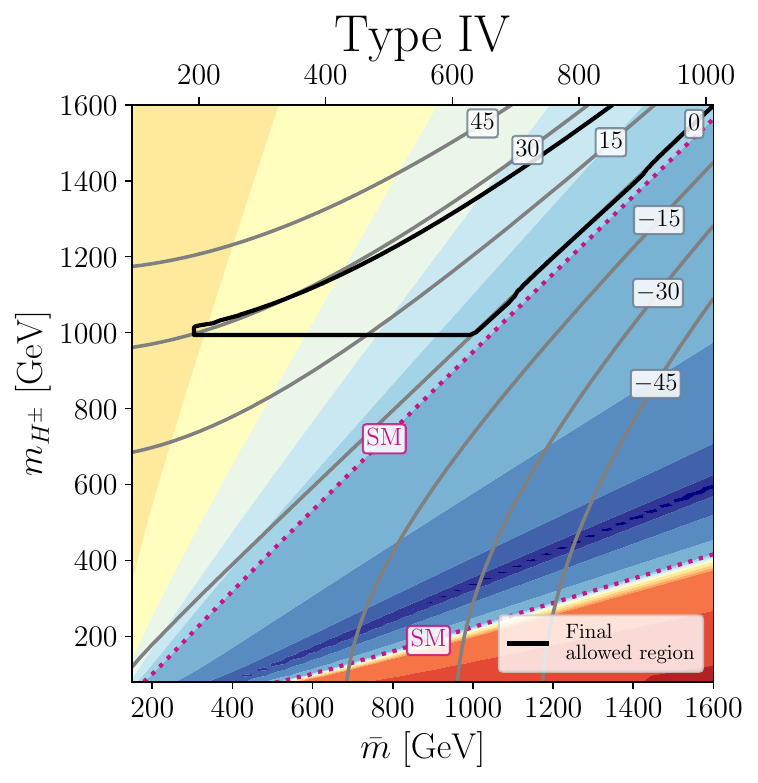}\includegraphics[width=0.54\textwidth]{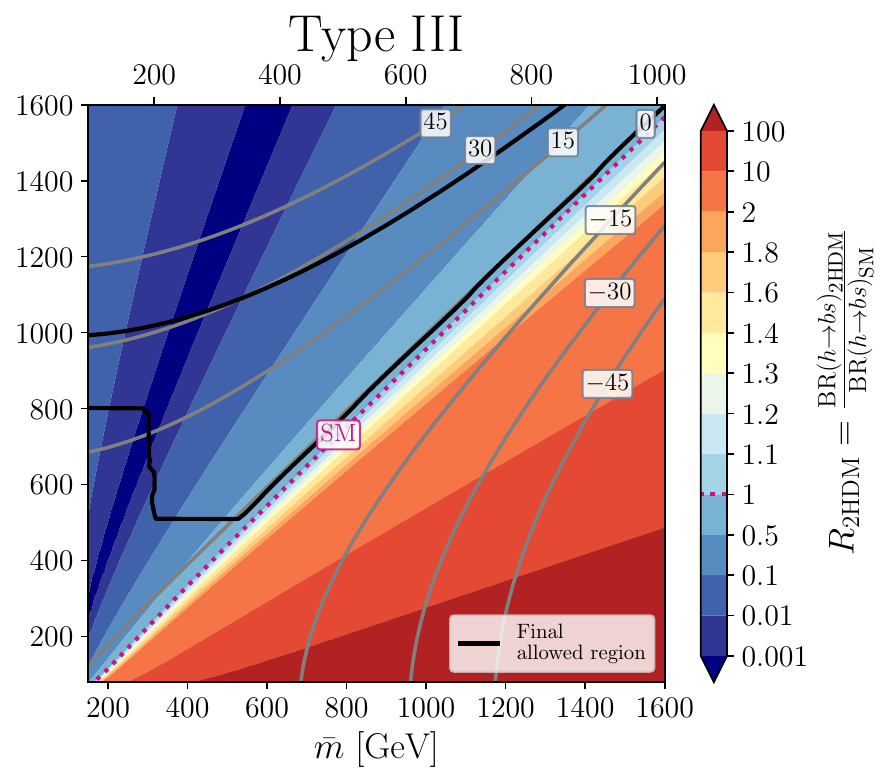}
\caption{Prediction of $\R=\br\left(h\to bs\right)_\mathrm{2HDM}/\br\left(h\to bs\right)_\mathrm{SM}$ in the plane $\bar m$-$\MHp$ with $\Mh = 125 \gev$, $\MH=\bar m + 50\gev$,  $\MA=\MHp$,  $\tan\beta=2$ and $\CBA=0$ for the four 2HDM types. 
The black contour lines indicates the boundary of the allowed region by all considered constraints.  
The dashed pink contour correspond to $\R=1$.
Solid grey lines shows contour lines with different values of $\lahHpHm$.}
\label{fig:mHpvsmbar}
\end{figure}

In \reffi{fig:mHpvsmbar} we show the prediction for $\R$ in the plane $\MHp$-$\bar m$ in the alignment limit for $\tb=2$, $\Mh = 125 \gev$, $\MH=\bar m - 50\gev$ and $\MA=\MHp$.
The $\R$ prediction for types I/IV are shown in the left column of the figure, while the 2HDM types II/III predictions  are shown in the right column.
In the upper axis we show the corresponding value of $m_{12}$,  which is related to $\bar m$ by the equation  $m_{12}^2=\bar m^2 \sin\beta\cos\beta=\frac{2}{5}\bar m^2$, for the chosen value of $\tb$.
Since we are in the alignment limit,  our prediction for the total width of $h$ does not depend on the model type and therefore the values of $\br\left(h \to bs\right)$ are equal in types I/IV and in types II/III. 

First, we comment on the shape of the allowed area by our set of constraints in the four types, whose boundaries are shown as solid black lines.
For all types, there is a diagonal line, close to $\MHp=\bar m$, coming from the potential stability constraints.
On the other hand, the boundary close to the $\lahHpHm=30$ contour corresponds to tree-level unitarity.
These theoretical constraints are model independent and therefore equal for the four types.
The lower bound for $\MHp$ in all types comes from $B\to X_s\gamma$ for types I/IV and from $B_s\to\mu\mu$ for types II/III.
Finally, there are some disallowed areas in types II/III for low values of $\bar m$ due to BSM Higgs-boson searches.
In both types, the bound below $\MHp\sim800\gev$ originates from a BSM search in the $pp\to H,\,A\to\tau\tau$ channel~\cite{ATLAS:2020zms}, while the additional bound in type III for $\MHp\gsim800\gev$ comes from a $gg\to A\to HZ\to bbll$ search~\cite{ATLAS:2018oht}.

Now, we turn to the discussion on the values for $\R$ in \reffi{fig:mHpvsmbar}.
Overall,  the contour lines for $\R$  follow straight lines with different slopes in the $\MHp$-$\bar m$ plane.
This reflects the dependence of the diagram~6 amplitude on these parameters, which is the dominant contribution to the process, as we have discussed above.
Furthermore,  the same interference pattern as before can be observed:
in types I/IV, for $\lahHpHm\gsim 0 \left(\lsim 0 \right) $ the interference is constructive (destructive), and the opposite happens in types II/III.
Since negative values for $\lahHpHm$ are disallowed by the theoretical constraints \cite{Arco:2022xum,Arco:2020ucn},  inside the allowed region we find an enhancement w.r.t.\ the SM prediction in types I/IV ($\R>1$) and a decrease w.r.t.\ the SM ($\R<1$) in types II/III.
 In particular, in this plane for types I/IV, $\R$ can reach values up to 1.4 close to the left ``tip" of the allowed region, around $\MHp\sim1\tev$ and $\bar m\sim300\gev$ (or $m_{12}\sim200\gev$).  
In types II/III, $\R$ can be orders of magnitude smaller than~1, just around that same region.
 In principle,  there should be a contour line where $\R$ is exactly zero,  but numerically that is not possible, and we find values as low as $\R\sim10^{-9}$, inside and outside the allowed region.
As a consequence of this destructive interference pattern, it is very difficult to find values for $\br\left(h\to bs\right)$ larger than the SM prediction within the allowed region in types II/III.

\begin{figure}[t!]
\centering
\includegraphics[width=0.4545\textwidth]{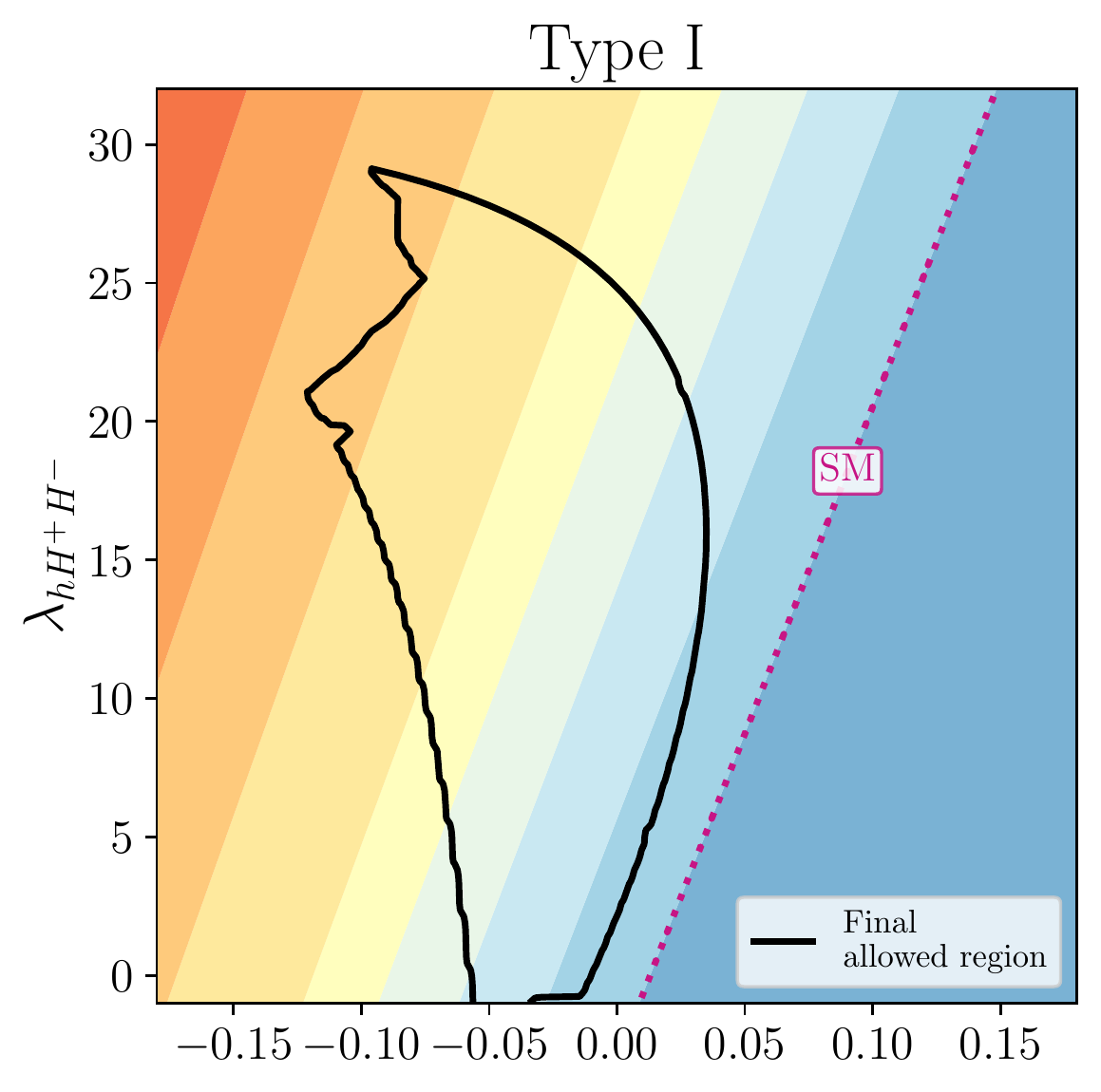}\includegraphics[width=0.56\textwidth]{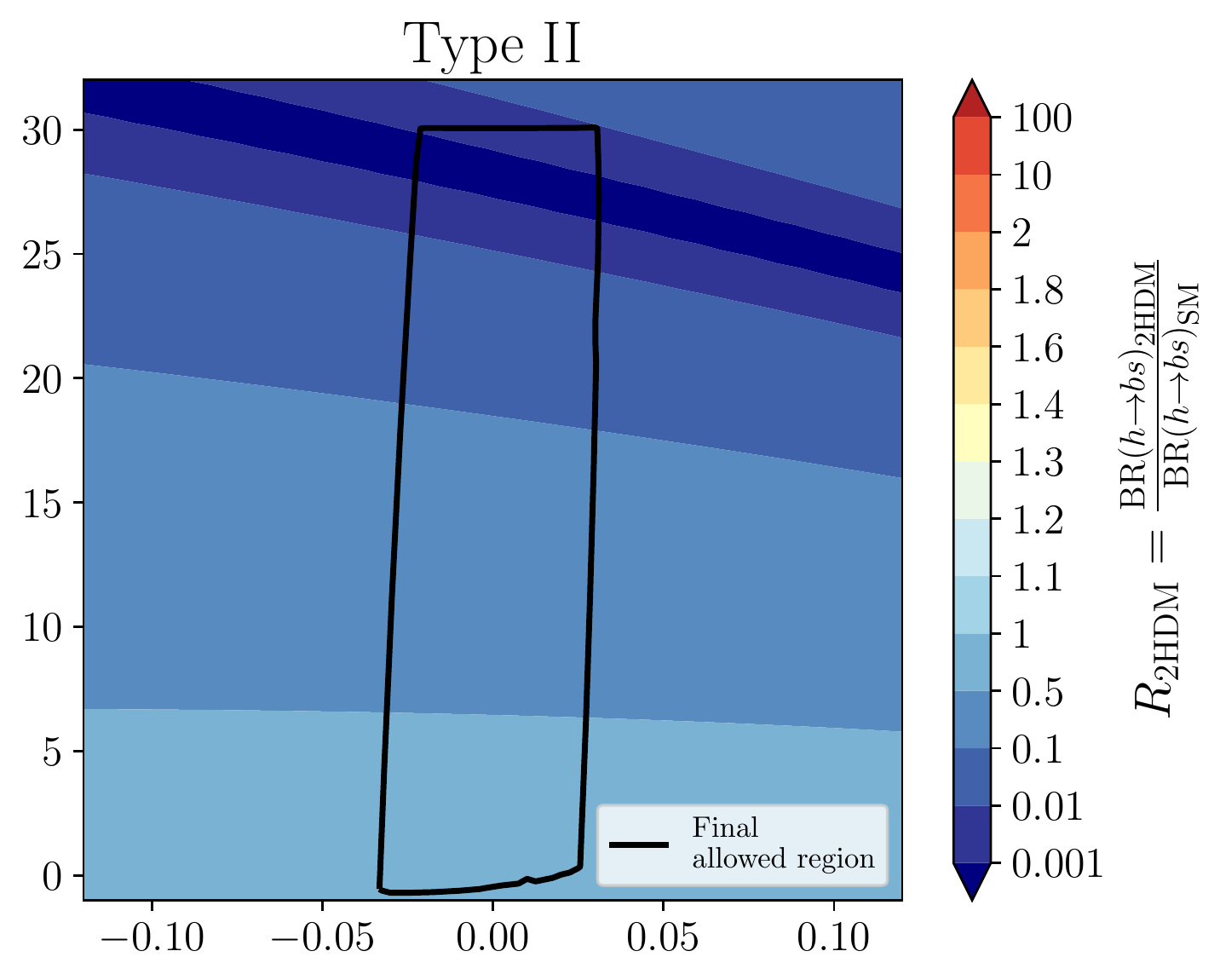} 
\includegraphics[width=0.4545\textwidth]{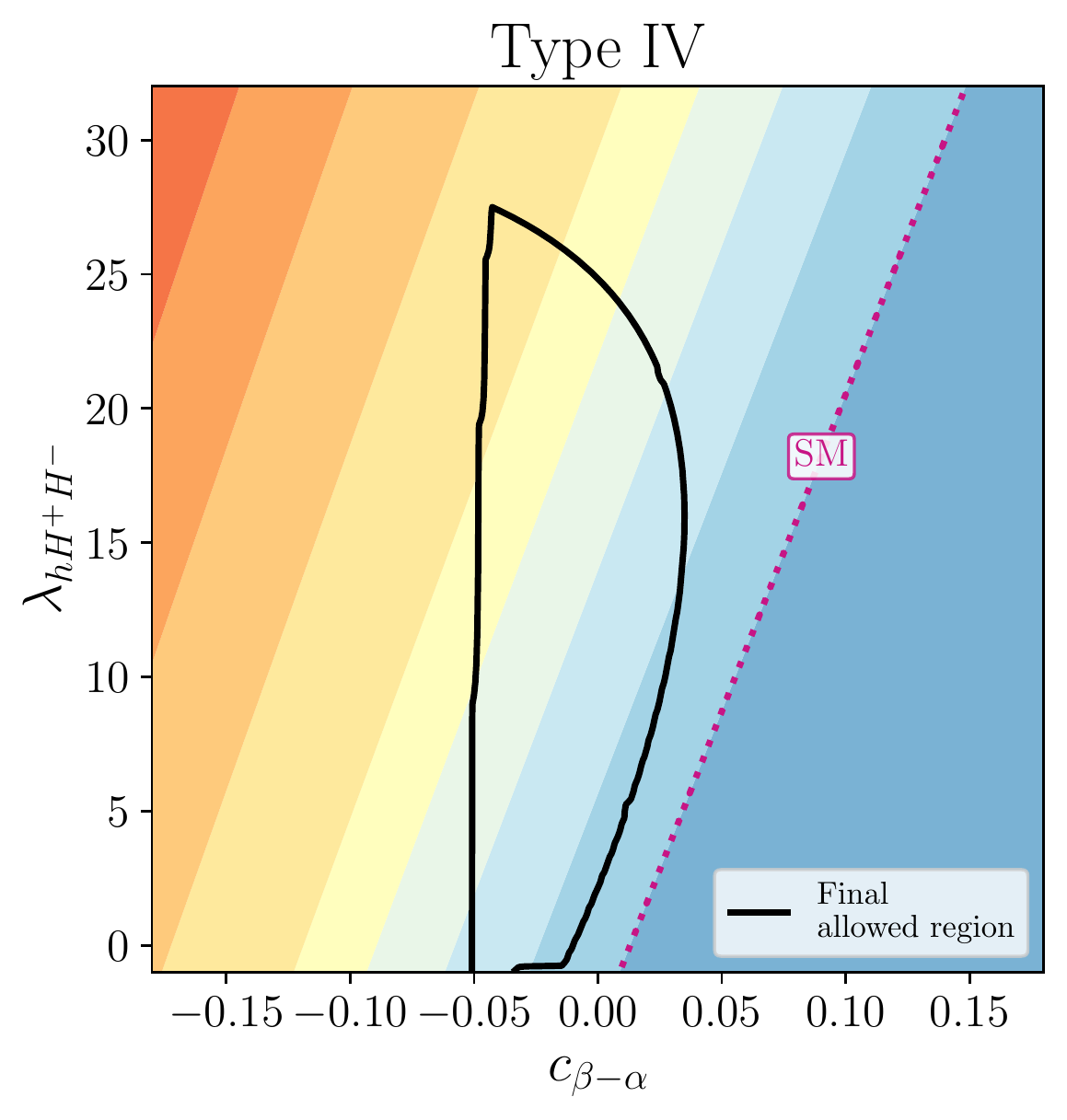}\includegraphics[width=0.56\textwidth]{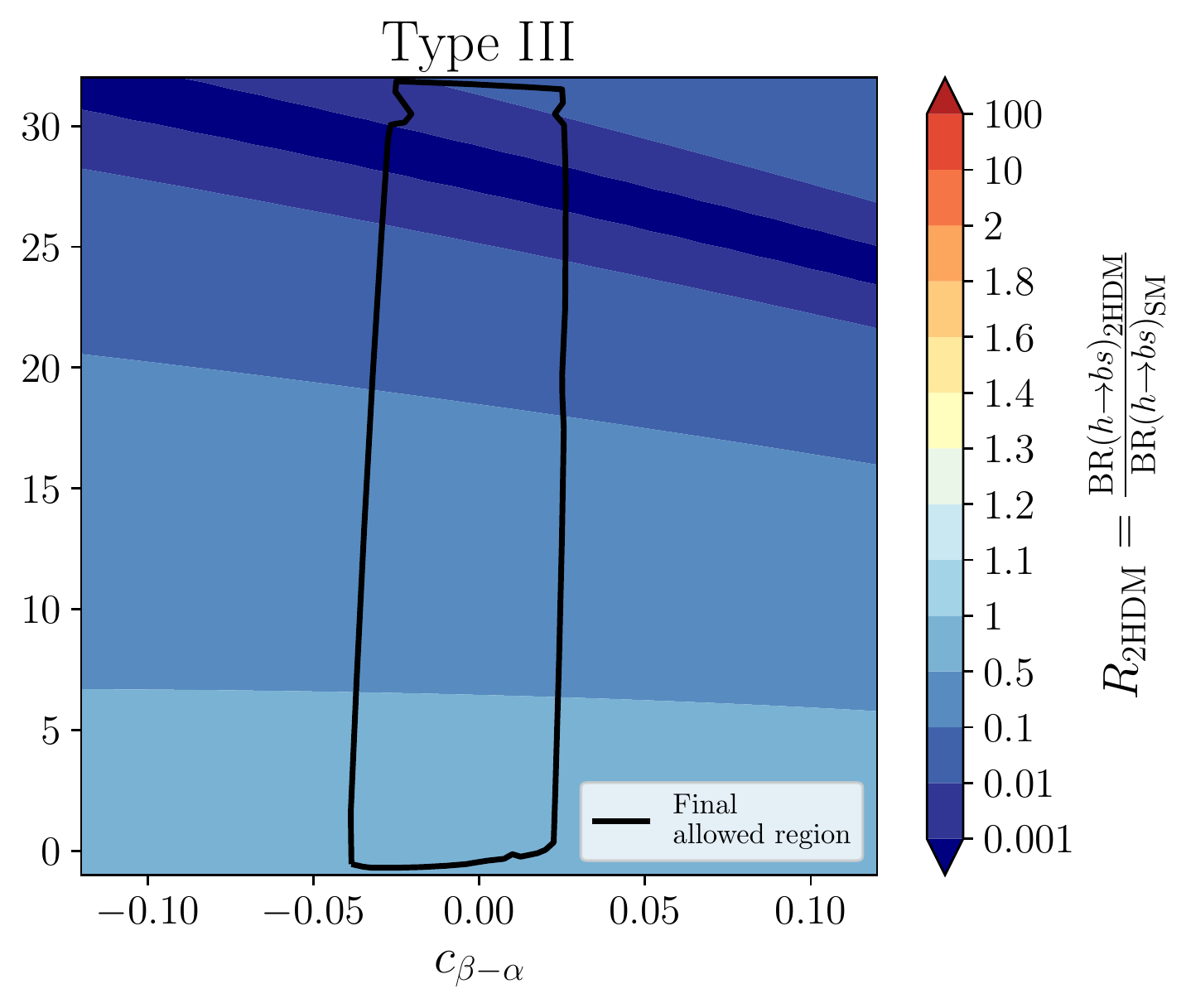} 
\caption{Prediction of $\R$~in the plane $\lahHpHm$-$\CBA$ with $\Mh = 125 \gev$, $\MHp=1000\gev$,  $\tan\beta=2$ and $\MH=\bar m + 120\gev$ for 2HDM types I and IV and $\MH=\bar m + 50\gev$ for 2HDM types II and III.  
The black contour lines indicates the boundary of the allowed region by all considered constraints.
The dashed pink contour correspond to $\R=1$. }
\label{fig:lambdavscbma}
\end{figure}

To emphasize on the relevance of the triple Higgs coupling $\lahHpHm$, in the prediction of $\br\left(h\to bs\right)$,  in \reffi{fig:lambdavscbma} we show $\R$ in the plane $\CBA$-$\lahHpHm$ for $\tb=2$ and $\MA=\MHp=1\tev$. 
Since in this figure,  in contrast to our previous plots, we are using as an input parameter $\lahHpHm$, according to \refeq{eq:lambda}, $\bar m$ and $m_{12}$ should be set to the respective derived values:
\begin{equation}
\bar m^2 = \frac{m_{12}^2}{\sin\beta\cos\beta} = \frac{\left(m_h^2+2\MHp^2\right)\SBA+2m_h^2\CBA\cot2\beta-v^2\lahHpHm}{2\SBA+2\CBA\cot2\beta}.
\end{equation}
Furthermore,  we choose $\MH=\bar m +120\gev$ for types I/IV, and $\MH=\bar m+50\gev$ for types II/III.
These particular choices are made with the goal to obtain the maximum possible distortion for $\R\neq1$, together with a sizable allowed region, specially in types I/IV.

First, we comment on the shape of the allowed regions plotted in \reffi{fig:lambdavscbma}.
The right curved boundary in types I/IV is present due to tree-level unitarity constraints.
The curved right border in type I covering values of $\lahHpHm$ from~0 to~20 is due to the potential stability constraints.
The additional upper irregular bounds for $\lahHpHm>20$ in type~I originate from an analysis combining several $pp\to H\to WW,\,ZZ$ searches~\cite{ATLAS:2018sbw}.
The ``left'' and ``right'' limits on $\CBA\sim\pm0.05$ in types II/III and also the ``left'' limit in type IV are given by the LHC rate measurements of the $125 \gev$ Higgs boson.
The upper bound for $\lahHpHm\sim30$ in type II come from a search in the $pp\to H\to\tau\tau$ channel~\cite{ATLAS:2020zms}.
In the case of type III,  this bound is not relevant in that region and the upper limit on $\lahHpHm$ appears due to the tree-level unitarity conditions. 
In addition, the small cracks in the allowed region present in type III around $\lahHpHm\sim30$ come from searches in the channel $pp\to H\to ZZ\to4l,\,2l2q,\,2l2\nu$ with $l=e,\,\mu$~\cite{CMS:2018amk}.
In all types, the bound for low values of $\lahHpHm$ are present due to the potential stability conditions.

Now we turn to the predictions for $\R$ in types I/IV, shown in the left panels of \reffi{fig:lambdavscbma}. 
For these types, the contour lines for this variable follow straight lines with a constant slope,  where $\R$ is larger for lower values of $\CBA$ and larger values of $\lahHpHm$.
This is in agreement with our discussion in \refse{sec:Gamma-noalign}, where negative values of $\CBA$ slightly enhance the prediction of $\br\left(h\to bs\right)$ because of the effects from the SM-like diagrams.
In addition,  the total decay width of $h$ has also a small effect on $\R$  in that direction because it decreases with $\CBA$ in types I/IV.
The predictions of $\R$ look very similar for both types, but there is a small,  hardly visible difference in the total width of $h$, and therefore in $\R$,  due to the differences in the leptonic decays.
Within the allowed region,  the maximum values of $\R$ are reached in both types for large (but allowed) values of $\lahHpHm\sim20-30$ and the lowest allowed value of $\CBA$.
Concretely,  $\R$ can reach values up to 1.7 in type I and 1.5 in type IV.
The choice for $\MH$ in types I/IV is such that the region allowed by the theoretical constraints is ``shifted'' to negative values of $\CBA$, where $\br\left(h\to bs\right)$ is larger.
Larger values for $\R$ are difficult to obtain while satisfying all considered constraints.
For instance,  the prediction for $\R$ could be larger for smaller values of $\tb$, but that would imply heavier $H^\pm$ values to avoid flavor constraints. 
Such large values of $\MHp$,  together with large values of $\lahHpHm$, easily violate the constraints from tree-level unitarity, especially outside the alignment limit.

Finally,  we comment on $\R$ in types II/III,  shown in the right column of \reffi{fig:lambdavscbma}. Both Yukawa types yield very similar predictions.
Again, there is a tiny difference in $\R$ between both types due to different $h$ decay rates to leptons. 
In this case,  the predictions for $\R$ decrease for large values of $\lahHpHm$, reaching a minimum in the upper region of the plot. 
Again, the reason is the destructive interference governed by diagram 6 in the prediction of $\Gamma\left(h\to bs\right)$.
The contour lines are nearly $\CBA$-independent, but the tilt is slightly stronger for larger values of $\lahHpHm$.
These extremely low values of the $\br\left(h\to bs\right)$ prediction can also be found in other regions of the parameter space of types II and III, in contrast to the maximum values found in types I and IV, which are only found in the indicated
parameter space.


\subsection{Experimental prospects for \boldmath{$h \to bs$}}
\label{sec:Prospects}

In this section we discuss on the future prospects to detect experimentally the decay channel $h\to bs$.
It is very challenging for the LHC, as well as the future HL-LHC, to measure any flavor-changing decays in the quark sector for the SM-like Higgs boson.
These searches suffer from extremely large hadronic backgrounds and therefore,  in practice,  they are inaccessible for these machines, see for instance \citere{Blankenburg:2012ex}. 
However, possible future $e^+e^-$ colliders provide a much cleaner environment and {\it{a priori}} could be able to detect this process. 
In particular, there is an analysis of the experimental prospects at the International Linear Collider (ILC)~\cite{Barducci:2017ioq}. It was found that the ILC could set an upper bound on  $\br\left(h\to bs\right)$ of order $10^{-3}$ at the 95\% CL.
To our knowledge,  there are no further analyses of this type for other lepton colliders, such as CLIC, FCC-ee or CEPC.
The possible future upper bound of \order{10^{-3}} is five orders of magnitude above the SM prediction.
Furthermore, in the previous section, we found that only enhancements up to 70\% or 50\% w.r.t.\ the SM (i.e. $\R=1.7$ and $1.5$) could be realized within 2HDM types I and IV, respectively.
In the case of 2HDM types II and III,  the predictions are in general below the SM prediction in almost the entire parameter space.
In consequence,  it appears that there is no collider projected at the medium-large term with sufficient experimental reach to measure such values of $\br\left(h\to bs\right)$, neither within the SM, nor in the 2HDM. 
Conversely, any signal detected in the $h \to bs$ channel at the projected future experiments would clearly point to models beyond the SM {\it and} the 2HDM with a softly broken $Z_2$ symmetry.

\medskip
One could ask the question if the prediction for $\R$ could be larger under some circumstances.
The main limiting factor to have a larger decay width for $h\to bs$ are the flavor constraints. 
In particular, for types I and IV,  $B\to X_s\gamma$ sets the strongest bounds at low $\tb$ in our analysis.
If in the future this bound relaxes (due to a shift in the central experimental value or any other reason),  the 2HDM could allow larger rates for  $h\to bs$.
For instance, from \reffi{fig:mHpvstb} it can be seen that $\Gamma\left(h\to bs\right)\sim10^{-9}\gev$ for $\tb=1$ and $\MHp=1\tev$,  leading to $\R\sim2.5$. 
Nevertheless,  it does not appear likely that such bounds would relax below $\tb=1$.
On the other hand,  large enhancements for very light $H^\pm$ could also be realized coming from diagrams without $\lahHpHm$.
However, these low values of $\MHp$ are strongly disfavored by experimental searches, see, e.g. \citere{ALEPH:2013htx} (LEP) and \citeres{ATLAS:2018gfm,CMS:2019bfg,CMS:2020osd} (LHC). 
Consequently, only a slightly larger prediction for $\R\sim3$ could be possible if the flavor bounds on the low $\tb$ region become less constraining in the future.
However, even such enhancements are still very far away from the expected experimental reach for this process.
Therefore, the conclusion holds that if any evidence of the decay $h\to bs$ is detected in the near future,  this would imply the presence of new physics beyond the SM {\it and} the $\cp$ conserving 2HDM with a softly broken $Z_2$ symmetry.

\medskip

We would like to comment briefly on the case that the discovered 125 GeV Higgs boson is identified with the heavy $\cp$-even boson $H$, implying that $\Mh < 125$ GeV and that alignment limit corresponds to $\CBA\to1$. 
This is known in the literature as ``inverted" hierarchy ($\Mh < \MH\sim125$ GeV), compared to the usual ``normal" hierarchy ($\Mh\sim 125$ GeV $<\MH$). 
We do not expect any significant impact on our conclusions for $\br(H \to bs)$ in the ``inverted" scenario compared to $\br(h \to bs)$ in the ``normal" one. 
In the alignment limit, the prediction of the decay rates for $H \to bs$ in the ``inverse" hierarchy is identical to $h \to bs$ in the ``normal" hierarchy but with $\MH \to \Mh$. 
Outside the alignment limit, there are some contributions in $H \to bs$ that change sign with respect to $h \to bs$ (like the vertices $hH^\pm G^\mp$ vs. $HH^\pm G^\mp$), but all of these would be small since they are proportional to $\SBA \sim 0$. The similarity between $h \to bs$ and $H \to bs$ in the ``normal" and ``inverted" hierarchies respectively is also observed in \citere{Arhrib:2004xu}. 
Furthermore, the fact that both $\cp$-even Higgs bosons are light in the ``inverted" scenario would imply an upper bound on the mass of $\MHp$ below 1 TeV (see, for instance, \citere{Bernon:2015qea}). 
Therefore, we expect very similar conclusions in both scenarios. 
In addition, one could also think that the ``inverted" scenario could favor lower values for $\MHp$, but as discussed above, such lower values of $\MHp$ are currently very discouraged by direct LEP/LHC searches. 

\medskip

Finally, we would like to highlight the differences between our work and the previous results of \citere{Arhrib:2004xu} from 2004. 
As discussed before, we update and complement the results of \citere{Arhrib:2004xu}, which investigated the $\br(h \to bs)$ only in the 2HDM types I and II. 
In the present work, we have analyzed the allowed values for $\br(h \to bs)$ in the four 2HDM types considering all the latest relevant constraints, including the measured properties of the later discovered Higgs boson with a mass around 125 GeV. 
Additionally, we also highlight the relevant role of the triple Higgs coupling $\lahHpHm$ in the final prediction of $\br(h \to bs)$. 
We discuss how this observable can exhibit non-decoupling effects induced by $\lahHpHm$, and we have also computed an effective vertex $hbs$ valid for large values of $\MHp$ in the alignment limit.


\section{Conclusions}
\label{sec:conclusions}

Within the 2HDM we have analyzed the phenomenology of the flavor-changing decays of the
SM-like Higgs boson to a strange-bottom quark pair, $h\to b\bar s$ and $h\to \bar bs$, which we denote together as $h\to bs$.
We have assumed the lightest $\cp$-even Higgs boson, denoted as $h$, to be the SM-like Higgs discovered by the LHC.
We consider a $Z_2$ symmetry that forbids FCNCs at the tree level, which is softly broken by the parameter $m_{12}$.
This is in contrast to other recent works that study this process within
a 2HDM with tree-level flavor-changing Yukawa
interactions \cite{Botella:2014ska,Crivellin:2017upt,Herrero-Garcia:2019mcy}, which can predict more sizable rates, but has to be fine-tuned to avoid large FCNCs disallowed by the current experiments.
Hence, this work provides an updated analysis of the previous result for
  one-loop generated FC Higgs decays within the 2HDM from
\citere{Arhrib:2004xu}, published before the discovery of the Higgs boson
  at $\sim 125 \gev$.
Furthermore, a significant focus of this paper is the analysis of the 2HDM contributions mediated by the charged Higgs bosons $H^\pm$. 
More concretely, we emphasize the role of the triple Higgs coupling
$\lahHpHm$, which was found to be crucial in this decay process.

We started our analysis of the $h\to bs$ decay width in the alignment limit,
which implies that the 2HDM
couplings of the $h$ that are present
in the SM recover their SM values.
Under this assumption, we found that the amplitude for the decay process $h\to bs$ can be split into two parts. 
The first part corresponds to the set of diagrams as in the SM, while
the second one consists of a pure 2HDM set of diagrams involving the charged Higgs boson, $H^\pm$. 
Among these 2HDM diagrams, the one with the triple Higgs coupling  $\lahHpHm$ stands out because its contribution can be of a similar order or even larger than the SM-like contribution. 
Consequently, the interference between the diagram with $\lahHpHm$ and all other contributions turns out to be
crucial in the analysis
  of the prediction of the decay width.
Concretely, this interference is positive (negative) for $\lahHpHm>0$ in the 2HDM types~I and~IV (types~II and~III).
Furthermore, this diagram can possibly lead to non-decoupling effects that do not vanish even for heavy $H^\pm$ due to the dependence on $\MHp$ of the triple Higgs coupling $\lahHpHm$.
We captured these one-loop effects of a heavy charged Higgs boson in the
prediction of $\Gamma(h \to bs)$ by a series expansion in inverse powers of
$\MHp$ up to $(m_{\rm EW}/\MHp)^2$, assuming this mass to be much larger than all other occurring masses.
Using this series expansion, we presented a one-loop effective vertex description of the heavy charged Higgs effects in $h\to bs$ in the alignment limit. 
This effective vertex approximation yields reasonably accurate predictions of the decay width for large values of $\MHp$ above 500 GeV,  and are  particularly good for low values of $\tan\beta<10$.

Subsequently, we analyzed the effects outside the alignment limit, obtaining similar conclusions since the effect of moderate values of $\CBA\neq0$ slightly modifies the $h$ couplings to other particles.
Therefore, the interference pattern between the diagram with $\lahHpHm$
and the remaining contributions holds similarly as in the alignment limit.

We studied the possible range of $\br(h\to
  bs)$ allowed by all relevant theoretical and current experimental
constraints on the 2HDM parameter space, based on previous analysis from \citeres{Arco:2022xum, Arco:2020ucn}.
Mainly positive values of $\lahHpHm$ are allowed by the 2HDM potential stability constraint.
Consequently, due to the interference patterns found in the decay width
prediction for the 2HDM Yukawa types, we find enhancements of
$\br\left(h\to bs\right)$ w.r.t.\ the SM prediction in types~I and~IV and a decrease in the allowed values for the BR in types~II and~III.
More concretely, in types~I and~IV, one can find enhancements of
the BR prediction of around 70\% and 50\%  w.r.t.\ the SM, respectively.
These predictions are realized for negative values of $\CBA$ and large values of $\lahHpHm$. 
The principal constraints that yield this maximum in
the BR predictions are light Higgs-boson rate measurements at the LHC,
which do not allow for a large deviation from the alignment limit, as well as
the theoretical constraints on the potential, due to the large values of $\lahHpHm$.
Regarding types~II and~III, the negative interference effects in these types
lead to significantly suppressed $\br\left(h\to bs\right)$ by up to several orders of magnitude compared to the SM.

Finally, we discussed the prospects to observe and probe the $h\to bs$ decay experimentally.
We conclude that the allowed ranges for the 2HDM prediction of $\br\left(h\to bs\right)$ are several orders of magnitude beyond the experimental reach of present and even future collider experiments.
Therefore, in the hypothetical case that any signal from this $h$ decay
channel is detected experimentally, it would require from physics beyond the SM {\it and} the 2HDM  with a softly broken $Z_2$ symmetry.


\subsection*{Acknowledgments}
\begingroup 
The present work has received financial support from the grant IFT Centro de Excelencia Severo Ochoa CEX2020-001007-S
  funded by MCIN/AEI/10.13039/501100011033.
The work of F.A.\ and S.H.\ was also supported in part by the
grant PID2019-110058GB-C21 funded by
MCIN/AEI/10.13039/501100011033 and by "ERDF A way of making Europe".
F.A.\ and M.J.H.\  also acknowledge financial support from the Spanish
``Agencia Estatal de Investigaci\'on'' (AEI) and the EU ``Fondo Europeo de
Desarrollo Regional'' (FEDER) 
through the project PID2019-108892RB-I00 funded by MCIN/AEI/10.13039/501100011033
and from the European Union's Horizon 2020 research and innovation
programme under the Marie Sklodowska-Curie grant agreement
No 860881-HIDDeN.
The work of F.A.\ was also supported by the Spanish Ministry of Science
and Innovation via an FPU grant with code FPU18/06634. 
\endgroup


\appendix
\section{ Feynman rules and expressions for the amplitudes of \boldmath{$h\to bs$}  in the 2HDM}
\label{app:amplitudes}

The 2HDM Feynman rules that are needed to obtain the amplitudes to the process $h\to bs$ are the following:
\begin{align}
hu\bar{u}:&\quad	-\frac{im_{u}\left(s_{\beta-\alpha}+\xi_{u}c_{\beta-\alpha}\right)}{v}, \\
hd\bar{d}:	&\quad-\frac{im_{d}\left(s_{\beta-\alpha}+\xi_{d}c_{\beta-\alpha}\right)}{v},\\
u\bar{d}H^{-}:&\quad	\frac{-i\sqrt{2}}{v}V_{ud}^{\ast}\left(m_{d}\xi_{d}P_{L}-m_{u}\xi_{u}P_{R}\right),\\
\bar{u}dH^{+}:&\quad	\frac{-i\sqrt{2}}{v}V_{ud}\left(m_{d}\xi_{d}P_{R}-m_{u}\xi_{u}P_{L}\right),\\
hG^{\pm}W^{\mp}:&\quad	\mp\frac{1}{2}igs_{\beta-\alpha}\left(p_{h}^{\mu}-p_{G}^{\mu}\right),\\
hH^{\pm}W^{\mp}:&\quad	\mp\frac{1}{2}igc_{\beta-\alpha}\left(p_{h}^{\mu}-p_{H^{\pm}}^{\mu}\right),\\
hG^{\pm}G^{\mp}:&\quad	-\frac{im_{h}^{2}s_{\beta-\alpha}}{v},\\
hH^{\pm}G^{\mp}:&\quad	-\frac{i\left(m_{h}^{2}-m_{H^{\pm}}^{2}\right)c_{\beta-\alpha}}{v},\\
hW^\pm W^\mp:&\quad igm_Ws_{\beta-\alpha}g^{\mu\nu}, \\
hH^{\pm}H^{\mp}:&\quad	-iv\lambda_{hH^{+}H^{-}},
\end{align}
with $\lahHpHm$ defined in \refeq{eq:lambda} and all momenta going inwards.

With the above Feynman rules,  the specific expressions for the amplitudes of each diagram with the same numbering and notation as in \reffi{fig:diagrams} and \refeq{eq:amp} are given by:
\begin{multline}
\mathcal{A}_1=m_q^2 \left(s_{\beta -\alpha }+c_{\beta -\alpha } \xi _u\right) \left\{m_s P_R \left[B_0^{\text{(1)}}+m_q^2 \left(C_0^{\text{(1)}}-2
   C_1^{\text{(1)}}\right)+m_h^2 C_1^{\text{(1)}} \right. \right. \\
   \left. \left.  +m_b^2 \left(C_0^{\text{(1)}}+2 \left(C_1^{\text{(1)}}+C_2^{\text{(1)}}\right)\right)\right]+m_b P_L
   \left[B_0^{\text{(1)}}+m_h^2 C_1^{\text{(1)}}+m_b^2 C_2^{\text{(1)}} \right. \right. \\
   \left. \left.  -m_s^2 \left(C_0^{\text{(1)}}+2 C_1^{\text{(1)}}+C_2^{\text{(1)}}\right)+m_q^2
   \left(3 C_0^{\text{(1)}}+2 \left(C_1^{\text{(1)}}+C_2^{\text{(1)}}\right)\right)\right]\right\}, 
\end{multline}
\begin{multline}
\mathcal{A}_2=-m_q^2 \left(s_{\beta -\alpha }+c_{\beta -\alpha } \xi _u\right)  \left\{ m_s P_R\left[ m_b^2 \xi _d^2 \left(C_0^{\text{(3)}}+2 C_1^{\text{(3)}}\right)+m_q^2 \xi
   _u^2 \left(C_0^{\text{(3)}}+2 C_2^{\text{(3)}}\right) \right. \right. \\
  \left. \left.  -\xi _d \xi _u \left(B_0^{\text{(3)}}+2 m_q^2 C_0^{\text{(3)}}+m_h^2 C_2^{\text{(3)}}\right) \right] +  m_b P_L \left[ m_q^2 \xi _u^2 \left(C_0^{\text{(3)}}+2 C_1^{\text{(3)}}\right)+m_s^2 \xi _d^2 \left(C_0^{\text{(3)}}+2 C_2^{\text{(3)}}\right) \right. \right. \\
  \left. \left.  -\xi _d \xi _u \left(B_0^{\text{(3)}}+2 m_q^2 C_0^{\text{(3)}}+m_h^2 C_2^{\text{(3)}}+m_s^2\left( C_0^{\text{(3)}}+ C_1^{\text{(3)}}+ C_2^{\text{(3)}}\right)  \right. \right. \right.  \\
  \left.  \left.  \left. -m_b^2 \left(C_0^{\text{(3)}}+C_1^{\text{(3)}}+C_2^{\text{(3)}}\right)\right) \right] \right\},  
\end{multline}
\begin{multline}  
\mathcal{A}_3= m_h^2 s_{\beta -\alpha }\left\{   m_s P_R \left[m_b^2 C_1^{\text{(2)}}+m_q^2 \left(C_0^{\text{(2)}}+C_2^{\text{(2)}}\right)\right]   \right.  \\
  \left.  + m_b  P_L \left[m_q^2 \left(C_0^{\text{(2)}}+C_1^{\text{(2)}}\right)+m_s^2 C_2^{\text{(2)}}\right] \right\},   
\end{multline}
\begin{multline}
\mathcal{A}_4=  c_{\beta -\alpha } \left(m_h^2-m_{H^{\pm }}^2\right)\left\{ m_s P_R \left[m_b^2 \xi _d C_2^{\text{(4)}}+m_q^2 \left(\xi _d C_0^{\text{(4)}}-\xi _u  \left(C_0^{\text{(4)}}+C_1^{\text{(4)}}+C_2^{\text{(4)}}\right)\right)\right] \right. \\
 \left.  - m_b P_L \left[\xi _d m_s^2 \left(C_0^{\text{(4)}}+C_1^{\text{(4)}}+C_2^{\text{(4)}}\right)-m_q^2 \xi _u \left(C_0^{\text{(4)}}+C_2^{\text{(4)}}\right)\right]\right\},   
\end{multline}
\begin{multline}
\mathcal{A}_5= -c_{\beta -\alpha }  \left(m_h^2-m_{H^{\pm }}^2\right) \left\{m_s P_R  \left[-m_q^2 \xi _u \left(C_0^{\text{(5)}}+C_2^{\text{(5)}}\right)+m_b^2 \xi _d  \left(C_0^{\text{(5)} }+C_1^{\text{(5)}}+C_2^{\text{(5)}}\right)\right] \right. \\
  \left.   - m_b P_L  \left[m_s^2 \xi _d C_2^{\text{(5)}}+m_q^2  \left(\xi _d C_0^{\text{(5)}}-\xi _u  \left(C_0^{\text{(5)}}+C_1^{\text{(5)}}+C_2^{\text{(5)}}\right)\right)\right] \right\} ,
\end{multline}
\begin{multline}
\mathcal{A}_6=-v^2  \lambda _{h H^- H^+}\left\{m_s P_R \left[-m_q^2 \xi _u \left(\xi _d C_0^{\text{(6)}}+\xi _u C_1^{\text{(6)}}\right)+m_b^2 \xi _d^2   \left(C_0^{\text{(6)}}+C_1^{\text{(6)}}+C_2^{\text{(6)}}\right)\right] \right. \\
  \left.  - m_b P_L \left[m_s^2 \xi _d^2   C_1^{\text{(6)}}+m_q^2 \xi _u \left(\xi _d C_0^{\text{(6)}}-\xi _u \left(C_0^{\text{(6)}}+C_1^{\text{(6)}}+C_2^{\text{(6)}}\right)\right)\right]\right\}   ,
\label{eq:A6}
\end{multline}
\begin{multline}
\mathcal{A}_7= -2 m_W^2 m_q^2 \left(s_{\beta -\alpha }+c_{\beta -\alpha } \xi _u\right)  \left\{m_s P_R \left[C_0^{\text{(1)}}+2 C_1^{\text{(1)}}\right]-m_b P_L  \left[C_0^{\text{(1)}}+2 \left(C_1^{\text{(1)}}+C_2^{\text{(1)}}\right)\right]\right\} ,
\end{multline}
\begin{multline}
\mathcal{A}_8=-m_W^2s_{\beta -\alpha }\left\{m_s P_R \left[m_q^2 \left(C_0^{\text{(2)}}-C_2^{\text{(2)}}\right)+m_b^2 \left(C_1^{\text{(2)}}+2 C_2^{\text{(2)}}\right)\right]  \right. \\
  \left.   - m_b P_L \left[B_0^{\text{(1)}}-\left(m_b^2-2 m_q^2-m_W^2\right) C_0^{\text{(2)}}+m_q^2 C_1^{\text{(2)}}+\left(m_b^2-m_h^2 \right)C_2^{\text{(2)}}\right]\right\} ,  
\end{multline}
\begin{multline}
\mathcal{A}_9= m_W^2 c_{\beta -\alpha }  \left\{ m_s P_R \left[m_b^2 \xi _d \left(C_0^{\text{(5)}}+C_1^{\text{(5)}}-C_2^{\text{(5)}}\right)-m_q^2 \xi _u   \left(C_0^{\text{(5)}}-C_2^{\text{(5)}}\right)\right]  \right. \\
  \left.  -m_b P_L \left[m_q^2 \xi _u  \left(-C_0^{\text{(5)}}+C_1^{\text{(5)}}+C_2^{\text{(5)}}\right)  \right. \right.  \\
  \left.  \left. -\xi _d \left(B_0^{\text{(1)}}+m_{H^{\pm }}^2 C_0^{\text{(5)}}-m_h^2 C_2^{\text{(5)}}-m_b^2 \left(C_0^{\text{(5)}}-C_2^{\text{(5)}}\right)\right)\right] \right\},
\end{multline}
\begin{multline}
\mathcal{A}_{10}=m_W^2 s_{\beta -\alpha } \left\{ m_s P_R \left[B_0^{\text{(1)}} + m_s^2 \left(2 C_1^{\text{(2)}}+C_2^{\text{(2)}}\right)+ m_q^2\left(2 C_0^{\text{(2)}} + C_2^{\text{(2)}} \right) +m_W^2 C_0^{\text{(2)}}  \right. \right. \\
  \left. \left.  +m_h^2 \left(-2 C_1^{\text{(2)}}+C_2^{\text{(2)}}\right)-m_b^2
   \left(C_0^{\text{(2)}}+C_1^{\text{(2)}}+C_2^{\text{(2)}}\right) \right] \right. \\
  \left.  + m_b P_L \left[ m_q^2 \left(-C_0^{\text{(2)}}+C_1^{\text{(2)}}\right)- m_s^2  \left(2
   C_1^{\text{(2)}}+C_2^{\text{(2)}}\right) \right] \right\},
\end{multline}
\begin{multline}
\mathcal{A}_{11}= m_W^2 c_{\beta -\alpha } \left\{m_s P_R \left[m_q^2 \xi _u   \left(C_0^{\text{(4)}}-C_1^{\text{(4)}}-C_2^{\text{(4)}}\right)-m_s^2 \xi _d \left(C_0^{\text{(4)}}+C_1^{\text{(4)}}-C_2^{\text{(4)}}\right)\right. \right. \\
   \left. \left.  +\xi _d   \left(B_0^{\text{(3)}}+m_W^2 C_0^{\text{(4)}}+m_b^2 C_1^{\text{(4)}}-m_h^2 \left(C_0^{\text{(4)}}+C_1^{\text{(4)}}+3   C_2^{\text{(4)}}\right)\right)\right] \right. \\
   \left.  + m_b P_L\left[m_s^2 \xi _d \left(C_0^{\text{(4)}}+C_1^{\text{(4)}}-C_2^{\text{(4)}}\right)-m_q^2 \xi _u   \left(C_0^{\text{(4)}}-C_2^{\text{(4)}}\right)\right]\right\},  
\end{multline}
\begin{multline}
\mathcal{A}_{12}= 4 m_W^4 s_{\beta -\alpha } \left\{m_s P_R C_2^{\text{(2)}} + m_b P_L C_1^{\text{(2)}}\right\},  \ \ \ \ \ \ \ \ \ \ \ \ \ \ \ \ \ \ \ \ \ \ \ \
\end{multline}
\begin{multline}
\mathcal{A}_{13}= -\frac{  s_{\beta -\alpha }+c_{\beta -\alpha } \xi _d}{ m_b^2-m_s^2 } \left\{m_b^2 m_s P_R  \left[m_s^2
   B_1^{\text{(1)}}+m_q^2 \left(2 B_0^{\text{(1)}}+B_1^{\text{(1)}}\right)\right]  \right. \\
  \left.   + m_b P_L \left[m_q^2 \left(m_b^2+m_s^2\right) B_0^{\text{(1)}}+\left(m_b^2+m_q^2\right) m_s^2
   B_1^{\text{(1)}}\right]\right\}, 
\end{multline}
\begin{multline}
\mathcal{A}_{14}=\frac{  s_{\beta -\alpha }+c_{\beta -\alpha } \xi _d}{m_b^2-m_s^2} \left\{m_b^2 m_s P_R \left[m_s^2 \xi _d^2  \left(B_0^{\text{(3)}}+B_1^{\text{(3)}}\right)  \right. \right.  \\
  \left.  \left.  +m_q^2 \xi _u \left(-2 \xi _d B_0^{\text{(3)}}+\xi _u
   \left(B_0^{\text{(3)}}+B_1^{\text{(3)}}\right)\right)\right]  + m_b P_L \left[m_q^2 m_s^2 \xi _u \left(\left(-\xi _d+\xi _u\right) B_0^{\text{(3)}}+\xi _u B_1^{\text{(3)}}\right)  \right. \right.  \\
  \left.  \left.  +m_b^2 \xi _d \left(-m_q^2 \xi _u B_0^{\text{(3)}}+m_s^2 \xi _d \left(B_0^{\text{(3)}}+B_1^{\text{(3)}}\right)\right)\right]\right\},   
\end{multline}
\begin{multline}
\mathcal{A}_{15} = - \frac{2 m_W^2 \left(s_{\beta -\alpha }+c_{\beta -\alpha } \xi _d\right)  }{  m_b^2-m_s^2 }\left\{m_b m_s \left(m_s P_L+m_b P_R\right)B_1^{\text{(1)}} \right\} ,   \ \ \ \ \ \ \ \ \ \ \ \ \ \ \ \ \ \ \ \ \ \ \ \ \ \ \ \ \ \ \ \ \
\end{multline}
\begin{multline}
\mathcal{A}_{16} = \frac{s_{\beta -\alpha }+c_{\beta -\alpha } \xi _d}{m_b^2-m_s^2}\left\{m_s P_R \left[m_q^2 \left(m_b^2+m_s^2\right) B_0^{\text{(2)}}+m_b^2 \left(m_q^2+m_s^2\right)  B_1^{\text{(2)}}\right] \right. \\
  \left.   +m_b m_s^2 P_L \left[m_b^2
   B_1^{\text{(2)}}+ m_q^2 \left(2 B_0^{\text{(2)}}+B_1^{\text{(2)}}\right)\right]\right\} ,  
\end{multline}
\begin{multline}
\mathcal{A}_{17} =  -\frac{s_{\beta -\alpha }+c_{\beta -\alpha } \xi _d}{ m_b^2-m_s^2 }\left\{ m_s   P_R \left[-m_q^2 m_s^2 \xi _d \xi _u B_0^{\text{(4)}} \right. \right.  \\
  \left.  \left.  +m_b^2 \left(m_s^2 \xi _d^2   \left(B_0^{\text{(4)}}+B_1^{\text{(4)}}\right)+m_q^2 \xi _u \left(\left(-\xi _d+\xi _u\right) B_0^{\text{(4)}}+\xi _u   B_1^{\text{(4)}}\right)\right)\right]  \right. \\
  \left. + m_b m_s^2 P_L \left[m_b^2 \xi _d^2 \left(B_0^{\text{(4)}}+B_1^{\text{(4)}}\right)+m_q^2 \xi _u   \left(-2 \xi _d B_0^{\text{(4)}}+\xi _u \left(B_0^{\text{(4)}}+B_1^{\text{(4)}}\right)\right)\right] \right\},  
\end{multline}
\begin{multline}
\mathcal{A}_{18} = \frac{2 m_W^2 \left(s_{\beta -\alpha }+c_{\beta -\alpha } \xi _d\right)}{m_b^2-m_s^2} \left\{ m_b m_s \left(m_s P_L+m_b P_R\right) B_1^{\text{(2)}} \right\},   \ \ \ \ \ \ \ \ \ \ \ \ \ \ \ \ \ \ \ \ \ \ \ \ \ \ \ \ \ \ \ \ \
\end{multline}
where $B_{i}$ and $C_{i}$ are Passarino-Veltman funtions with the following arguments: $ B_{i}^{(1)}=B_{i}(m_{s}^{2},m_{q}^{2},m_{W}^{2})$, $ B_{i}^{(2)}=B_{i}(m_{b}^{2},m_{q}^{2},m_{W}^{2})$,  $B_{i}^{(3)}=B_{i}(m_{s}^{2},m_{H^{\pm}}^{2},m_{q}^{2})$,  $B_{i}^{(4)}=B_{i}(m_{b}^{2},m_{H^{\pm}}^{2},m_{q}^{2})$, $ C_{i}^{(1)}=C_{i}(m_{h}^{2},m_{s}^{2},m_{b}^{2},m_{q}^{2},m_{q}^{2},m_{W}^{2})$, $ C_{i}^{(2)}=C_{i}(m_{b}^{2},m_{h}^{2},m_{s}^{2},m_{q}^{2},m_{W}^{2},m_{W}^{2})$,  $C_{i}^{(3)}=C_{i}(m_{b}^{2},m_{h}^{2},m_{s}^{2},m_{H^{\pm}}^{2},m_{q}^{2},m_{q}^{2})$, $ C_{i}^{(4)}=C_{i}(m_{s}^{2},m_{b}^{2},m_{h}^{2},m_{H^{\pm}}^{2},m_{q}^{2},m_{W}^{2})$,  $C_{i}^{(5)}=C_{i}(m_{b}^{2},m_{s}^{2},m_{h}^{2},m_{H^{\pm}}^{2},m_{q}^{2},m_{W}^{2})$ and $ C_{i}^{(6)}=C_{i}(m_{h}^{2},m_{s}^{2},m_{b}^{2},m_{H^{\pm}}^{2},m_{H^{\pm}}^{2},m_{q}^{2}) $, where we use the same convention for the Passarino-Veltman function as in {\tt FormCalc} \cite{Hahn:1998yk}.

If we consider the approximation that the first and second quark generation are massless we get the following reduced expressions:
\begin{multline}
\mathcal{A}_{1} =  m_b m_t^2 P_L \left(s_{\beta -\alpha }+c_{\beta -\alpha } \xi _u\right) \left\{B_0^{\text{(1)}}+m_h^2 C_1^{\text{(1)}}+m_b^2
   C_2^{\text{(1)}} \right.  \\
  \left.  +m_t^2 \left(3 C_0^{\text{(1)}}+2 \left(C_1^{\text{(1)}}+C_2^{\text{(1)}}\right)\right)\right\},
\end{multline}
\begin{multline}
\mathcal{A}_{2} = -m_b m_t^2 P_L \xi
   _u \left(s_{\beta -\alpha }+c_{\beta -\alpha } \xi _u\right)\left\{m_t^2 \xi _u \left(C_0^{\text{(3)}}+2 C_1^{\text{(3)}}\right)  \right.  \\
  \left.  +\xi _d
   \left(-B_0^{\text{(3)}}-2 m_t^2 C_0^{\text{(3)}}-m_h^2 C_2^{\text{(3)}}+m_b^2
   \left(C_0^{\text{(3)}}+C_1^{\text{(3)}}+C_2^{\text{(3)}}\right)\right)\right\},
\end{multline}
\begin{multline}
\mathcal{A}_{3} = m_b m_h^2 m_t^2 P_L s_{\beta -\alpha }
   \left\{C_0^{\text{(2)}}+C_1^{\text{(2)}}\right\},
\ \ \ \ \ \ \ \ \ \ \ \ \ \ \ \ \ \ \ \ \ \ \ \ \ \ \ \ \ \ \ \ \ \ \ \ \ \ \ \ \ \ \ \ \ \ \ \ \ \
\end{multline}
\begin{multline}
\mathcal{A}_{4} = c_{\beta -\alpha } m_b m_t^2 \left(m_h^2-m_{H^{\pm }}^2\right) P_L \xi _u
   \left\{C_0^{\text{(4)}}+C_2^{\text{(4)}}\right\},
\ \ \ \ \ \ \ \ \ \ \ \ \ \ \ \ \ \ \ \ \ \ \ \ \ \ \ \ \ \ \ \ \ \ \ \ \ \ \
\end{multline}
\begin{multline}
\mathcal{A}_{5} = c_{\beta -\alpha } m_b m_t^2 \left(m_h^2-m_{H^{\pm }}^2\right) P_L \left\{\xi _d
   C_0^{\text{(5)}}-\xi _u \left(C_0^{\text{(5)}}+C_1^{\text{(5)}}+C_2^{\text{(5)}}\right)\right\},
\ \ \ \ \ \ \ \ \ \ \ \ \ \ 
\end{multline}
\begin{multline}
\mathcal{A}_{6} = -m_b m_t^2 P_L v^2\lambda _{h H^- H^+}
   \xi _u \left\{-\xi _d C_0^{\text{(6)}}+\xi _u \left(C_0^{\text{(6)}}+C_1^{\text{(6)}}+C_2^{\text{(6)}}\right)\right\},
\ \ \ \ \ \ \ \ \ \ \ \ \ \ \ \ \ \ \ \ \ 
\end{multline}
\begin{multline}
\mathcal{A}_{7} = 2 m_b m_W^2 m_t^2 P_L
   \left(s_{\beta -\alpha }+c_{\beta -\alpha } \xi _u\right)
    \left\{C_0^{\text{(1)}}+2 \left(C_1^{\text{(1)}}+C_2^{\text{(1)}}\right)\right\},
\ \ \ \ \ \ \ \ \ \ \ \ \ \ \ \ \ \ \ \ \ \ \ \
\end{multline}
\begin{multline}
\mathcal{A}_{8} = m_b m_W^2 P_L s_{\beta -\alpha } \left\{B_0^{\text{(1)}}+\left(2 m_t^2+m_W^2\right) C_0^{\text{(2)}}+m_t^2 C_1^{\text{(2)}}-m_h^2
   C_2^{\text{(2)}} \right.  \\
  \left.  +m_b^2 \left(-C_0^{\text{(2)}}+C_2^{\text{(2)}}\right)\right\},
\end{multline}
\begin{multline}
\mathcal{A}_{9} = c_{\beta -\alpha } m_b m_W^2 P_L  \left\{m_t^2 \xi _u
   \left(C_0^{\text{(5)}}-C_1^{\text{(5)}}-C_2^{\text{(5)}}\right) \right.  \\
  \left.  +\xi _d \left(B_0^{\text{(1)}}+\left(-m_b^2+m_{H^{\pm }}^2\right)
   C_0^{\text{(5)}}+\left(m_b^2-m_h^2\right) C_2^{\text{(5)}}\right)\right\},
\end{multline}
\begin{multline}
\mathcal{A}_{10} = m_b m_W^2 m_t^2 P_L s_{\beta -\alpha }
   \left\{-C_0^{\text{(2)}}+C_1^{\text{(2)}}\right\},
\ \ \ \ \ \ \ \ \ \ \ \ \ \ \ \ \ \ \ \ \ \ \ \ \ \ \ \ \ \ \ \ \ \ \ \ \ \ \ \ \ \ \ \ \
\end{multline}
\begin{multline}
\mathcal{A}_{11} =  c_{\beta -\alpha } m_b m_W^2 m_t^2 P_L \xi _u 
   \left\{-C_0^{\text{(4)}}+C_2^{\text{(4)}}\right\},
\ \ \ \ \ \ \ \ \ \ \ \ \ \ \ \ \ \ \ \ \ \ \ \ \ \ \ \ \ \ \ \ \ \ \ \ \ \ \ \ \ \ \ \ \ \ \ \ 
\end{multline}
\begin{multline}
\mathcal{A}_{12} =4 m_b m_W^4 P_L s_{\beta -\alpha }  C_1^{\text{(2)}},
\ \ \ \ \ \ \ \ \ \ \ \ \ \ \ \ \ \ \ \ \ \ \ \ \ \ \ \ \ \ \ \ \ \ \ \ \ \ \ \ \ \ \ \ \ \ \ \ \ \ \ \ \ \ \ \ \ \ \ \
\end{multline}
\begin{multline}
\mathcal{A}_{13} = - m_b m_t^2 P_L
   \left(s_{\beta -\alpha }+c_{\beta -\alpha } \xi _d\right) B_0^{\text{(1)}},
\ \ \ \ \ \ \ \ \ \ \ \ \ \ \ \ \ \ \ \ \ \ \ \ \ \ \ \ \ \ \ \ \ \ \ \ \ \ \ \ \ \ \ \ \ \ 
\end{multline}
\begin{multline}
\mathcal{A}_{14} = - m_b m_t^2 P_L \xi _d \left(s_{\beta -\alpha }+c_{\beta
   -\alpha } \xi _d\right) \xi _u B_0^{\text{(3)}},
\ \ \ \ \ \ \ \ \ \ \ \ \ \ \ \ \ \ \ \ \ \ \ \ \ \ \ \ \ \ \ \ \ \ \ \ \ \ \ \ \ \ \ \ \ \ 
\end{multline}
\begin{multline}
\mathcal{A}_{15} = \mathcal{A}_{16} =  \mathcal{A}_{17} = \mathcal{A}_{18} = 0.
\ \ \ \ \ \ \ \ \ \ \ \ \ \ \ \ \ \ \ \ \ \ \ \ \ \ \ \ \ \ \ \ \ \ \ \ \ \ \ \ \ \ \ \ \ \ 
\end{multline}


\section{Large \boldmath{$\MHp$} series expansion}
\label{app:expansion}

In this appendix we present the analytical expressions of the relevant Passarino-Veltman functions 
needed to perform the large $\MHp$ series expansion described in \refse{sec:largeMHp}.
We consider only the $t$-mediated loop diagrams and we neglect the masses of the bottom and the strange quarks in the loop functions.

The Passarino-Veltman functions in diagrams 2, 14 and 17 have to be known up order one in $x_{H^\pm}\propto\MHp^{-2}$:
\begin{equation}
B_0^{(3)}=B_0^{(4)}=B_0\left(0,\MHp^2,m_t^2 \right) =  \frac{1}{\epsilon} + 1 +  \log \left(\frac{\mu ^2}{\MHp^2}\right)-\frac{m_t^2 \log \left(\frac{ \MHp^2}{ m_t^2}\right)}{\MHp^2},
\end{equation}
\begin{equation}
B_1^{(3)}=B_1^{(4)}=B_1\left(0,\MHp^2,m_t^2 \right) =- \frac{1}{2\epsilon} -\frac{3}{4} - \frac{1}{2} \log\left( \frac{\mu^2}{\MHp^2} \right)
+ \frac{ m_t^2 \left( \log \left(\frac{ \MHp^2}{ m_t^2}\right)-\frac{1}{2}\right)}{\MHp^2} ,
\end{equation}
\begin{equation}
C_0^{(3)} = C_0\left(0,\Mh^2,0,\MHp^2,m_t^2,m_t^2\right) = \frac{2 \sqrt{4 r-1} \arctan\left(\frac{1}{\sqrt{4 r-1}}\right)- \log \left(\frac{ \MHp^2}{ m_t^2}\right)-1}{\MHp^2} ,
\end{equation}
\begin{equation}
C_1^{(3)} = C_1\left(0,\Mh^2,0,\MHp^2,m_t^2,m_t^2\right)=  \frac{ -\sqrt{4 r-1} \arctan\left(\frac{1}{\sqrt{4 r-1}}\right)+\frac{1}{2}\log \left(\frac{ \MHp^2}{ m_t^2}\right)+\frac{1}{4} }{\MHp^2} ,
\end{equation}
\begin{align}
C_2^{(3)} =C_2^{(3)} &= C_2\left(0,\Mh^2,0,\MHp^2,m_t^2,m_t^2\right) 
\nonumber \\
 &= \frac{ 6 r-6 (r-1) \sqrt{4 r-1} \mathrm{\ arccot}\left(\sqrt{4 r-1}\right)-3 \log \left(\frac{ \MHp^2}{ m_t^2}\right)-1 }{9\MHp^2} ,
\end{align}
with $r=m_t^2/m_h^2$ and $\epsilon$ as defined below \refeq{eq:div}.

The Passarino-Veltman functions in diagram 6 have to be known up to order two (i.e. $x_{H^\pm}^2\propto\MHp^{-4}$), 
due to the presence of $\lahHpHm$ as a common multiplicative factor:
\begin{equation}
C_0^{(6)} = C_0\left(\Mh^2,0,0,\MHp^2,\MHp^2,m_t^2\right)  =- \frac{1}{\MHp^2} - \frac{\frac{m_h^2}{12}- m_t^2 \log \left(\frac{ \MHp^2}{ m_t^2}\right)+m_t^2}{\MHp^{4}} ,
\label{C0-diag6}
\end{equation}
\begin{equation}
C_1^{(6)} = C_1\left(\Mh^2,0,0,\MHp^2,\MHp^2,m_t^2\right) = \frac{1}{4\MHp^2} +  \frac{m_h^2-9 m_t^2}{36\MHp^{4}} ,
\label{C1-diag6}
\end{equation}
\begin{equation}
C_2^{(6)} = C_2\left(\Mh^2,0,0,\MHp^2,\MHp^2,m_t^2\right) = \frac{1}{2\MHp^2} +  \frac{m_h^2-36 m_t^2 \log \left(\frac{ \MHp^2}{ m_t^2}\right)+54 m_t^2}{36\MHp^{4}}.
\label{C2-diag6}
\end{equation}

The result of this series expansions were performed with the help of the code {\tt PackageX} \cite{Patel:2015tea}.  

The final contribution from diagrams 2, 6, 14 and 17 in the alignment limit up to
order one in $x_{H^\pm}$ is:
\begin{multline}
\mathcal{A}_{H^\pm}  =m_t^2\xi_u\left(\frac{1}{2}\xi_u-2\xi_d\right)\left(m_sP_R+m_bP_L\right)
+\frac{m_t^2}{6\MHp^2}\log \left(\frac{\MHp^2}{m_t^2}\right) \Bigl( 5 m_b^3 P_L \xi_d \xi_u
\\
+m_b P_L \xi_d \left(5 m_s^2 (2 \xi_d-\xi_u)-2 m_h^2 \xi_u\right) 
+2 m_s P_R \xi_u \left(5 m_t^2 \xi_u-m_h^2 \xi_d\right) \Bigr)
\\
+\frac{1 }{36 m_h^2 \MHp^2} \Bigl( 12 m_t^2 \sqrt{4r-1} \mathrm{\ arccot}\left(\sqrt{4r-1}\right) \left(m_b^3 P_L \xi_d \xi_u \left(2 m_t^2-5 m_h^2\right)
\right.  \\ \left. 
+m_b P_L \xi_d \left(2 m_h^4 \xi_u+5 m_h^2 \left(m_s^2 (\xi_u-2 \xi_d)+2 m_t^2 \xi_u\right)+2 m_s^2 m_t^2 (2 \xi_d-\xi_u)\right)
\right.  \\ \left. 
+2 m_s P_R \xi_u \left(m_h^4 \xi_d+5 m_h^2 m_t^2 (\xi_d-\xi_u)+2 m_t^4 \xi_u\right)\right)-m_t^2 \xi_d \xi_u \left(m_b P_L \left(m_h^2 \left(-31 m_b^2-72 \bar{m}^2
\right. \right. \right.  \\ \left. \left. \left. 
+46 m_h^2+31 m_s^2\right)+24 m_t^2 \left(m_b^2+5 m_h^2-m_s^2\right)\right)+2 m_h^2 m_s P_R \left(-36 \bar{m}^2+23 m_h^2+60 m_t^2\right)\right)
\\ 
+m_b m_s \xi_d^2 \left(-18 \bar{m}^2 m_h^2 (m_b P_R+m_s P_L)+11 m_h^4 (m_b P_R+m_s P_L)+26 m_h^2 m_s m_t^2 P_L-48 m_s m_t^4 P_L\right)
\\ 
+m_t^2 \xi_u^2 \left(m_b m_h^2 P_L \left(11 m_h^2-18 \bar{m}^2\right)+m_s P_R \left(-18 \bar{m}^2 m_h^2+11 m_h^4+26 m_h^2 m_t^2-48 m_t^4\right)\right)\Bigr),
\end{multline}
where we have neglected the $m_t$-independent terms, because they vanish due to the unitarity of the CKM matrix.

Under the assumption that $m_s=0$ we get:
\begin{multline}
\mathcal{A}_{H^\pm}  \simeq m_t^2m_bP_L\xi_u \biggl\{\frac{1}{2}\xi_u-2\xi_d 
 + \frac{1}{36 m_h^2\MHp^2}
 \biggl(m_h^2 \left(31 m_b^2+72 \bar{m}^2-46 m_h^2\right) \xi_d
 \\ 
 -24 \left(m_b^2+5 m_h^2\right) m_t^2 \xi_d
+m_h^2 \left(11 m_h^2-18 \bar{m}^2\right) \xi_u+6 \xi_d \biggl(2\sqrt{4r-1} \left(m_b^2 \left(2 m_t^2-5 m_h^2\right)
 \right. \\ \left. 
+2 \left(m_h^4+5 m_h^2 m_t^2\right)\right) \mathrm{arccot}\left(\sqrt{4r-1}\right)+m_h^2 \left(5 m_b^2-2 m_h^2\right) \log\left(\frac{\MHp^2}{m_t^2}\right) \biggr) \biggr) \biggl\} .
\end{multline}
Finally,  if $m_b$ is neglected with respect to $m_t$, $m_h$ and $m_{12}$ we get:
\begin{multline}
\mathcal{A}_{H^\pm}  \simeq m_t^2m_bP_L\xi_u \biggl\{ \frac{1}{2}\xi_u-2\xi_d
 + \frac{1}{36 \MHp^2} \biggl( 36 \bar{m}^2 \left(2 \xi_d-\frac{1}{2}\xi_u\right)
 +m_h^2 (11 \xi_u-46 \xi_d)
 \\ 
-120 m_t^2 \xi_d+12 \xi_d \left(2\left(m_h^2+5 m_t^2\right) \sqrt{4r-1} \mathrm{\ arccot}\left(\sqrt{4 r-1}\right) 
-m_h^2 \log \left(\frac{\MHp^2}{m_t^2 }\right)\right)\biggr) \biggr\} .
\end{multline}


\section{Numerical input parameters and considerations}
\label{app:num}
The correct implementation of the unitarity of the CKM matrix is key to compute the partial width of the decay $h\to bs$.
Therefore,  we parametrized the CKM matrix $V_\mathrm{CKM}$ in the ``standard parametrization'' consisting in three rotation angles and a  $\cp$ phase $\delta$,  yielding a CKM matrix that is unitary by construction.
Furthermore,  we considered the $\cp$ phase to be zero, since the $\cp$ effects in the observable $h\to bs$ are negligible.  
By doing this,  the CKM matrix becomes real and the unitarity conditions are easier to fulfill numerically.  
In summary, the CKM matrix that we used in our computation is:

\begin{equation}
V_{\mathrm{CKM}}=\left(\begin{array}{ccc}
c_{12}c_{13} & s_{12}c_{13} & s_{13}\\
-s_{12}c_{23}-c_{12}s_{23}s_{13} & c_{12}c_{23}-s_{12}s_{23}s_{13} & s_{23}c_{13}\\
s_{12}s_{23}-c_{12}c_{23}s_{13} & -c_{12}s_{23}-s_{12}c_{23}s_{13}& c_{23}c_{13}
\end{array}\right),
\label{eq:CKM}
\end{equation}
where $s_{12}=0.22500$, $s_{23}=0.04182$, $s_{13}=0.00369$, according to PDG \cite{ParticleDataGroup:2022pth}. 
 In order to ensure with high numerical accuracy the unitarity of the CKM matrix, we have computed
the Passarino-Veltman functions present in the amplitude (shown in the previous appendix) with the help of {\tt LoopTools} \cite{Hahn:1998yk} with quadruple precision.
 
 In addition, we also considered the masses of the final-state quarks, $m_b$ and $m_s$, in the $\overline{\text{MS}}$ scheme at an energy scale of the SM-like Higgs boson mass, in order to include leading contributions from QCD corrections. 
We used the values as given in~\citere{Huang:2020hdv}:  $m_{b}\left(m_h\right)=2.768\gev$ and $m_s\left(m_h\right)=0.052\gev$. The rest of the masses involved in the computation were taken from PDG~\cite{ParticleDataGroup:2022pth}:
 $m_u=2.16\times10^{-3}\gev$, $m_c=1.27\gev$, $m_t=172.5\gev$, $m_W=80.377\gev$ and $m_h=125.25\gev$.
In our calculation we employed the $G_F$-scheme, where the EW parameters are derived from the Fermi constant $G_F=1.166 378 8\times 10^{-5}\gev^{-2}$.  In particular,
$v=(\sqrt{2}G_F)^{-\frac{1}{2}}\simeq246.22\gev$ and 
 $g=2m_W/v\simeq0.65$.

\end{document}